%% file: main.tex
\documentclass[unsortedaddress,twocolumn,eqsecnum,aps,showkeys]{revtex4-1}  
\usepackage[dvips]{graphicx}
\usepackage{dcolumn}
\usepackage{longtable}
\usepackage{bm}
\usepackage{color}
\usepackage{amsmath}
\begin{document}
\input{title.tex}
\section{Introduction}
\label{sec:intro}
\input{intro.tex} 
\section{Formal Equations}
\label{sec:dressed}
\input{dressed.tex}
\section{Implementation}
\label{sec:implement}
\input{implement.tex}
\section{Computational Details}
\label{sec:details}
\input{details.tex}
\section{Results}
\label{sec:results}
\input{results.tex}
\section{Conclusion}
\label{sec:conclude}
\input{conclude.tex} 
\begin{acknowledgments}
\input{thanks.tex}

\end{acknowledgments}
\appendix
\section{Kohn-Sham-based Second-Order Polarization Propagator}
\label{app:soppa}
\input{soppa.tex}
\section{Tables of D-TDDFT and CISD excitation energies and oscillator strengths}
\label{app:tables}
\input{tab.tex}
\newpage
\bibliographystyle{myaip}
\bibliography{refs}
\end{document}

%% file: title.tex
\title{Assessment of Dressed Time-Dependent Density-Functional Theory for the Low-Lying Valence States of 28 Organic Chromophores}

\author{Miquel Huix-Rotllant$^1$}
\email[]{Miquel.Huix@UJF-Grenoble.Fr}
\author{Andrei Ipatov$^1$}
\author{Angel Rubio$^2$}
\author{Mark E. Casida$^1$}
\email[]{Mark.Casida@UJF-Grenoble.Fr}
\affiliation{
        $^1$ Laboratoire de Chimie Th\'eorique,
        D\'epartement de Chimie Mol\'ecularie (DCM, UMR CNRS/UJF 5250),
        Institut de Chimie Mol\'eculaire de Grenoble (ICMG, FR2607),
        Universit\'e Joseph Fourier (Grenoble I),
        301 rue de la Chimie, BP 53,
        F-38041 Grenoble Cedex 9, France\\
        $^2$ Nano-Bio Spectroscopy Group and ETSF Scientific Development Centre, 
        Departamento de F{\'{\i}}sica de Materiales, 
        Universidad del Pa{\'{\i}}s Vasco, E-20018 San Sebasti{\'{a}}n (Spain);\\
        Centro de F{\'{\i}}sica de Materiales 
        CSIC-UPV/EHU-MPC and DIPC, 
        E-20018 San Sebasti{\'{a}}n (Spain); \\
        Fritz-Haber-Institut der Max-Planck-Gesellschaft, 
        Faradayweg 4-6 , D-14195 Berlin-Dahlem, (Germany).}

  
\begin{abstract}

Almost all time-dependent density-functional theory (TDDFT) calculations of excited states make use of the adiabatic approximation, which implies a frequency-independent exchange-correlation kernel that limits applications to one-hole/one-particle states. To remedy this problem, Maitra \textit{et al.}[J.Chem.Phys. \textbf{120}, 5932 (2004)] proposed dressed TDDFT (D-TDDFT), which includes explicit two-hole/two-particle states by adding a frequency-dependent term to adiabatic TDDFT. This paper offers the first extensive test of D-TDDFT, and its ability to represent excitation energies in a general fashion. We present D-TDDFT excited states for 28 chromophores and compare them with the benchmark results of Schreiber \textit{et al.}[J.Chem.Phys. \textbf{128}, 134110 (2008).] We find the choice of functional used for the A-TDDFT step to be critical for positioning the 1h1p states with respect to the 2h2p states. We observe that D-TDDFT without HF exchange increases the error in excitations already underestimated by A-TDDFT. This problem is largely remedied by implementation of D-TDDFT including Hartree-Fock exchange.

\end{abstract}

\keywords{time-dependent density-functional theory, exchange-correlation kernel, adiabatic approximation, frequency dependence, many-body perturbation theory, excited states, organic chromophores}

\maketitle


%% file: intro.tex
Time-dependent density-functional theory (TDDFT) is a popular approach
for modeling the excited states of medium- and large-sized molecules.
It is a formally exact theory \cite{RG84}, which involves an exact
exchange-correlation (xc) kernel with a role similar to the
xc-functional of the Hohenberg-Kohn-Sham ground-state theory. Since
the exact xc-functional is not known, practical calculations involve
approximations.  Most TDDFT applications use the so-called adiabatic
approximation which supposes that the xc-potential responds
instantaneously and without memory to any change in the
self-consistent field \cite{RG84}. The adiabatic approximation limits
TDDFT to one hole-one particle (1h1p) excitations (i.e., single
excitations), albeit dressed to include electron correlation
effects \cite{C05}. Overcoming this limitation is desirable for
applications of TDDFT to systems in which 2h2p excitations (i.e.,
double excitations) are required, including the excited states of
polyenes, open-shell molecules, and many common photochemical
reactions \cite{SWS+06,R99,LKQ+06}. Maitra \textit{et
al.} \cite{MZC+04,CZM+04} proposed the dressed TDDFT (D-TDDFT) model,
an extension to adiabatic TDDFT (A-TDDFT) which explicitly includes
2h2p states.  The D-TDDFT kernel adds frequency-dependent terms from
many-body theory to the adiabatic xc-kernel.  While initial results on
polyenic systems appear encouraging \cite{CZM+04,MW09,MMR+10}, no
systematic assessment has been made for a large set of molecules.  The
present article reports the first systematic study of D-TDDFT for a
large test set namely, the low-lying excited states of 28 organic
molecules for which benchmark results exist \cite{SSS+08a,SSS+08b}.
This study has been carried out with several variations of D-TDDFT
implemented in a development version of the density-functional theory
(DFT) code {\sc deMon2k} \cite{deMon2k}.

The formal foundations of TDDFT were laid out by Runge and Gross
(RG) \cite{RG84} which put on rigorous grounds the earlier TDDFT
calculations of Zangwill and Soven \cite{ZS80}. The original RG
theorems showed some subtle problems \cite{V98}, which have been since
re-examined, criticized, and improved \cite{R96,M05,V08} providing a
remarkably well-founded theory (for a recent review see \cite{NM09}.)
A key feature of this formal theory is a time-dependent Kohn-Sham
equation containing a time-dependent xc-potential describing the
propagation of the density after a time-dependent perturbation is
applied to the system. Casida used linear response (LR) theory to
derive an equation for calculating excitation energies and oscillator
strengths from TDDFT \cite{C95}. The resultant equations are similar
to the random-phase approximation (RPA) \cite{BP53},
\begin{equation}
\label{eq:casidarpa}
  \left[ \begin{array}{cc} {\bf A}(\omega) & {\bf B}(\omega) \\ -{\bf
 B}^*(\omega) & -{\bf
 A}^*(\omega) \end{array} \right] \left[ \begin{array}{c} {\bf X}\\
 {\bf Y} \end{array} \right]= \omega\left[
\begin{array}{c}
   {\bf X} \\ {\bf Y}
\end{array}\right] \, .
\end{equation}
However ${\bf A}(\omega)$ and ${\bf B}(\omega)$ explicitly include the
Hartree (H) and xc kernels,
\begin{eqnarray}
  \label{eq:rpablocks} A_{ai\sigma,bj\tau} &=&
   (\epsilon^\sigma_a-\epsilon^\sigma_i)\delta_{ij}\delta_{ab}\delta_{\sigma\tau}
   + (ia|f_{Hxc}^{\sigma,\tau}(\omega)|bj) \nonumber\\
   B_{ai\sigma,bj\tau} &=& (ia|f_{Hxc}^{\sigma,\tau}(\omega)|jb) \, ,
\end{eqnarray}
where $\epsilon_p^\sigma$ is the KS orbital energy for spin $\sigma$,
and
\begin{eqnarray}
\label{eq:intdef}
(pq|&&f(\omega)|rs)= \\ &&\int{d^3r\int{d^3r' \phi^{*}_p({\bf
r})\phi_q({\bf r}) f({\bf r},{\bf r}';\omega)\phi^{*}_r({\bf
r}')\phi_s({\bf r}')}} \nonumber \, .
\end{eqnarray} 
Here and throughout this paper we use the following notation of
indexes: $i,j,...$ are occupied orbitals, $a,b,...$ are virtual
orbitals, and $p,q,...$ are orbitals of unspecified nature.

In chemical applications of TDDFT, the Tamm-Dancoff approximation
(TDA) \cite{HH99},
\begin{equation}
\label{eq:casidatda}
{\bf A}(\omega){\bf X} = \omega{\bf X} \, ,
\end{equation}
improves excited state potential energy surfaces \cite{CGG+00,CJI+07},
though sacrificing the Thomas-Reine-Kuhn sum rule.  Although the
standard RPA equations provide only 1h1p states, the exact LR-TDDFT
equations include also 2h2p states (and higher-order $n$h$n$p states)
through the $\omega$-dependence of the xc part of the kernel
$f^{\sigma,\tau}_{xc}(\omega)$. However, the matrices ${\bf
A}(\omega)$ and ${\bf B}(\omega)$ are supposed $\omega$-independent in
the adiabatic approximation to the xc-kernel , thereby losing the
non-linearity of the LR-TDDFT equations and the associated 2h2p (and
higher) states.

Double excitations are essential ingredients for a proper description
of several physical and chemical processes. Though they do not appear
directly in photo-absorption spectra, (i.e., they are dark states),
signatures of 2h2p states appear indirectly through mixing with 1h1p
states, thereby leading to the fracturing of main peaks into
satellites.  In open-shell molecules such mixing is often required in
order to maintain spin symmetry \cite{C05,CIC06,LL10}.  Perhaps more
importantly dark states often play an essential important role in
photochemistry and explicit inclusion of 2h2p states is often
considered necessary for a minimally correct description of conical
intersections \cite{LKQ+06}. A closely-related historical, but still
much studied, problem is the location of 2h2p states in
polyenes \cite{WHS+05,SWS+06,A10,CD88,SSN92,LC00,BBK+04,CP06,MTM08},
partly because of the importance of the polyene retinal in the
photochemistry of vision \cite{W72,G72,H72}.

It is thus manifest that some form of explicit inclusion of 2h2p
states is required within TDDFT when attacking certain types of
problems.  This has lead to various attempts to include 2h2p states in
TDDFT. One partial solution was given by spin-flip
TDDFT \cite{K01,WZ04} which describes some states which are 2h2p with
respect to the ground state by beginning with the lowest triplet state
and including spin-flip excitations \cite{WZ06,MG09,RVA10,SHK03}.
However, spin-flip TDDFT does not provide a general way to include
double excitations. Strengths and limitations of this theory have been
discussed in recent work \cite{HNI+10}.

The present article focuses on D-TDDFT, which offers a general model
for including explicitly 2h2p states in TDDFT. D-TDDFT was initially
proposed by Maitra, Zhang, Cave and Burke as an \textit{ad hoc}
many-body theory correction to TDDFT \cite{MZC+04}. They subsequently
tested it on butadiene and hexatriene with encouraging
results \cite{CZM+04}. The method was then reimplimented and tested on
longer polyenes and substituted polyenes by Mazur \textit{et
al.} \cite{MW09,MMR+10}.

In the present work, we consider several variants of D-TDDFT,
implement and test them on the set of molecules proposed by
Schreiber \textit{et al.} \cite{SSS+08a,SSS+08b} The set consists of
28 organic molecules whose excitation energies are well characterized
both experimentally or through high-quality \textit{ab initio}
wavefunction calculations.

This paper is organized as follows.  Section~\ref{sec:dressed}
describes D-TDDFT in some detail and the variations that we have
implemented.  Section~\ref{sec:implement} describes technical aspects
of how the formal equations were implemented in {\sc deMon2k}, as well
as additional features which were implemented specifically for this
study.  Section~\ref{sec:details} describes computational details such
as basis sets and choice of geometries.  Section~\ref{sec:results}
presents and discusses results. Finally, section~\ref{sec:conclude}
concludes.


%% file: dressed.tex
D-TDDFT may be understood as an approximation to exact equations for
the xc-kernel \cite{HC10}. This section reviews D-TDDFT and the
variations which have been implemented and tested in the present work.

An \textit{ab initio} expression for the xc-kernel may be derived from
many-body theory, either from the Bethe-Salpeter equation or from the
polarization propagator (PP) formalism \cite{RSB+09,C05}. Both
equations give the same xc-kernel,
\begin{eqnarray} 
\label{eq:fxcmbpt}
f_{xc}({\bf x}&,&{\bf x}';\omega) = \int {
 d^3x_{1} \int{d^3x_{2} \int{d^3x_{3} \int{d^3x_{4}}}}} \\
 &&\Lambda_s({\bf x};{\bf x}_1,{\bf x}_2;\omega) K({\bf x}_1, {\bf
 x}_2; {\bf x}_3 , {\bf x}_4;\omega) \Lambda^\dagger({\bf x}_3,{\bf
 x}_4;{\bf x}';\omega) \nonumber \, ,
\end{eqnarray}
where $x_p=({\bf r}_p,\sigma_p)$, $K$ is defined as
\begin{eqnarray}
\label{eq:soppakernel}
K({\bf x}_1,{\bf x}_2&;&{\bf x}_3,{\bf x}_4;\omega) = \\
&& \Pi^{-1}_s({\bf x}_1,{\bf x}_2;{\bf x}_3,{\bf x}_4;\omega)
-\Pi^{-1}({\bf x}_1,{\bf x}_2;{\bf x}_3,{\bf x}_4;\omega) \nonumber
\end{eqnarray}
and $\Pi$ and $\Pi_s$ are respectively the interacting and
non-interacting polarization propagators, which contribute to the pole
structure of the xc-kernel. The interacting and non-interacting
localizers, $\Lambda$ and $\Lambda_s$ respectively, convert the
4-point polarization propagators into the 2-point TDDFT quantities
(4-point and 2-point refer to the space coordinates of each kernel.)
The localization process introduces an extra $\omega$-dependence into
the xc-kernel.  Interestingly, Gonze and Scheffler \cite{GS99} noticed
that, when we substitute the interacting by the non-interacting
localizer in Eq.~(\ref{eq:fxcmbpt}), the localization effects can be
neglected for key matrix elements of the xc-kernel at certain
frequencies, meaning that the $\omega$-dependence exactly cancels the
spatial localization.  More importantly, removing the localizers
simply means replacing TDDFT with many-body theory terms.  To the
extent that both methods represent the same level of approximation,
excitation energies and oscillator strengths are unaffected, though
the components of the transition density will change in a finite basis
representation.  In Ref.~\cite{C05}, Casida proposed a PP form of
D-TDDFT without the localizer.  In Ref.~\cite{HC10}, Huix-Rotllant and
Casida gave explicit expressions for an \textit{ab initio}
$\omega$-dependent xc-kernel derived from a Kohn-Sham-based
second-order polarization propagator (SOPPA) formula.

The calculation of the xc-kernel in SOPPA can be cast in RPA-like
form.  In the TDA approximation, we obtain
\begin{equation}
  \label{eq:lowdin} \left[ {\bf A}_{11} + {\bf A}_{12} \left( \omega
  {\bf 1}_{22} - {\bf A}_{22} \right)^{-1} {\bf A}_{21} \right ] {\bf
  X} = \omega {\bf X} \, ,
\end{equation}
which provides a matrix representation of the second-order
approximation of the many-body theory kernel $K({\bf x}_1,{\bf
x}_2;{\bf x}_3,{\bf x}_4;\omega)$.  The blocks ${\bf A}_{11}$, ${\bf
A}_{21}$ and ${\bf A}_{22}$ couple respectively single excitations
among themselves, single excitations with double excitations and
double excitations among themselves.  In Appendix~\ref{app:soppa} we
give explicit equations for these blocks in the case of a SOPPA
calculation based on the KS Fock operator. We recall that in the SOPPA
kernel, the ${\bf A}_{11}$ is frequency independent, though it
contains some correlation effects due to the 2h2p states. All
$\omega$-dependence is in the second term and it originates from the
${\bf A}_{22}$ coupled to the ${\bf A}_{11}$ block.

The D-TDDFT kernel is a mixture of the many-body theory kernel and the
A-TDDFT kernel. This mixture was first defined by Maitra and
coworkers \cite{MZC+04}. They recognized that the single-single block
was already well represented by A-TDDFT, therefore substituting the
expression of ${\bf A}_{11}$ in Eq.~(\ref{eq:lowdin}) for the
adiabatic ${\bf A}$ block of Casida's equation
[Eq.(\ref{eq:rpablocks}).] This many-body theory and TDDFT mixture is
not uniquely defined.  As we will show, different combinations of
${\bf A}_{11}$ and ${\bf A}_{22}$ give rise to completely different
kernels, and not all combinations include correlation effects
consistently. In the present work, we wish to test several definitions
of the D-TDDFT kernel by varying the ${\bf A}_{11}$ and ${\bf A}_{22}$
blocks. For each D-TDDFT kernel, we will compare the excitation
energies against high-quality \textit{ab initio} benchmark
results. This will allow us to make a more accurate definition of the
D-TDDFT approach.

We will use two possible adiabatic xc-kernels in the ${\bf A}_{11}$
matrix: the pure LDA xc-kernel and a hybrid xc-kernel. Usually, hybrid
TDDFT calculations are based on a hybrid KS wavefunction. Our
implementations are done in {\sc deMon2k}, a DFT code which is limited
to pure xc-potentials in the ground-state calculation. Therefore, we
have devised a hybrid calculation that does not require a hybrid DFT
wavefunction. Specifically, the RPA blocks used in Casida's equations
are modified as
\begin{eqnarray}
  \label{eq:hybrid} A_{ai\sigma,bj\tau}
  &=& \left[ \epsilon^\sigma_a\delta_{ab} + c_0\cdot(a|{\hat
  M}_{xc}|b)\right]\delta_{ij}\delta_{\sigma\tau} \\
  &-&\left[ \epsilon^\sigma_i\delta_{ij} + c_0\cdot(i|{\hat
  M}_{xc}|j)\right]\delta_{ab}\delta_{\sigma\tau} \nonumber \\ &+&
  (ai|(1-c_0)\cdot f^{\sigma\tau}_x + c_0\cdot{\hat \Sigma}_x^{HF}+
  f^{\sigma\tau}_{Hc}|jb) \nonumber \\ B_{ai\sigma,bj\tau} &=&
  (ai|(1-c_0)\cdot f^{\sigma\tau}_x +
  c_0\cdot{\hat \Sigma}_x^{HF}+f^{\sigma\tau}_{Hc}|bj) \nonumber \, ,
\end{eqnarray} 
where ${\hat \Sigma}^{HF}_x$ is the HF exchange operator and ${\hat
M}_{xc}={\hat \Sigma}^{HF}_x-v_{xc}$ provides a first-order conversion
of KS into HF orbital energies. We note that the first-order
conversion is exact when the space of occupied KS orbitals coincides
with the space of occupied HF orbitals.  Also, the conversion from KS
to HF orbital energies introduces an effective particle number
discontinuity.

Along with the two definitions of the $A_{11}$ block, we will also
test different possible definitions for the ${\bf A}_{22}$
block. First, we will test a independent particle approximation (IPA)
estimate of ${\bf A}_{22}$, consisting of diagonal KS orbital energy
differences.  It was shown in Ref.~\cite{HC10} that such a block also
appears in a second-order \textit{ab initio} xc-kernel. We will call
that combination D-TDDFT.  Second, we will use a first-order
correction to the IPA estimate of ${\bf A}_{22}$. This might give an
improved description for the placement of double
excitations \cite{TJ95}. We call that combination extended D-TDDFT
(x-D-TDDFT).  We note that this is the approach of Maitra \textit{et
al.} \cite{MZC+04}.

\begin{table}
  \caption{Summary of the methods used in this work. CIS, CISD and
           A-TDDFT are the standard methods, whereas the (x-)D-CIS and
           (x-)D-TDDFT are the variations we use. The kernel $f_{Hxc}$
           represents the Hartree kernel plus the exchange-correlation
           kernel of DFT in the adiabatic approximation,
           $\Sigma_x^{HF}$ is the HF exchange and $\Delta\epsilon$ is
           a zeroth-order estimate for a double
           excitation.}  \begin{tabular}{cccc} \hline Method & ${\bf
           A}_{02}$ & ${\bf A}_{11}$ & ${\bf A}_{22}$ \\ \hline CIS &
           No & $f_H + \Sigma_x^{HF}$ & 0 \\ A-TDDFT & No & $f_{Hxc}$
           & 0 \\ CISD & Yes & $f_H + \Sigma_x^{HF}$ &
           $\Delta\epsilon^{HF}+$ first-order \\ D-CIS & No & $f_H
           + \Sigma_x^{HF}$ & $\Delta\epsilon^{KS}$ \\ x-D-CIS & No &
           $f_H + \Sigma_x^{HF}$ & $\Delta\epsilon^{KS}+$
           first-order \\ D-TDDFT & No & $f_{Hxc}$ &
           $\Delta\epsilon^{KS}$ \\ x-D-TDDFT & No & $f_{Hxc}$ &
           $\Delta\epsilon^{KS}+$ first-order \\ \hline \end{tabular}
\label{tab:methods}
\end{table}

In Table~\ref{tab:methods} we summarize the different variants of
D-TDDFT and D-CIS, according to ${\bf A}_{11}$ and ${\bf A}_{22}$
blocks. All the methods share the same ${\bf A}_{12}$ block unless the
${\bf A}_{22}$ block is 0, in which case the ${\bf A}_{12}$ is also
0. We recall that only the standard CISD has a coupling block ${\bf
A}_{01}$ and ${\bf A}_{02}$ with the ground state, but none of the
methods used in this paper has.


%% file: implement.tex
We have implemented the equations described in Sec.~\ref{sec:dressed}
in a development version of {\sc deMon2k}. The standard code now has a
LR-TDDFT module \cite{IFP+06}. In this section, we briefly detail the
necessary modifications to implement D-TDDFT.

{\sc deMon2k} is a Gaussian-type orbital DFT program which uses an
auxiliary basis set to expand the charge density, thereby eliminating
the need to calculate 4-center integrals. The implementation of TDDFT
in {\sc deMon2k} is described in Ref.~\cite{IFP+06}. Note that newer
versions of the code have abandoned the charge conservation constraint
for TDDFT calculations.  For the moment, only the adiabatic LDA (ALDA)
can be used as TDDFT xc-kernel.

Asymptotically-corrected (AC) xc-potentials are needed to correctly
describe excitations above the ionization threshold, which is placed
at minus the highest-occupied molecular orbital
energy \cite{CJC+98}. Such corrections are not yet present in the
master version of {\sc deMon2k}. Since such a correction was deemed
necessary for the present study, we have implemented Hirata \textit{et
al.}'s improved version
\cite{HZA+03} of Casida and Salahub's AC potential \cite{CS00} in our 
development version of {\sc deMon2k}.

\begin{figure}
 \caption{Necessary double excitations that need to be included in the
          truncated 2h2p space to maintain pure spin
          symmetry.}  \includegraphics[scale=0.7]{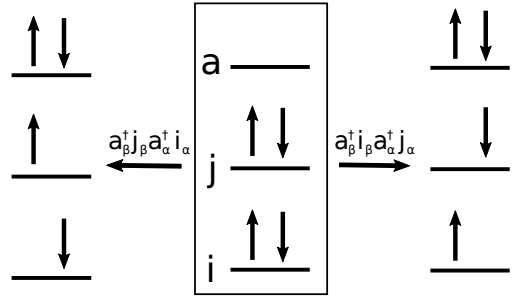} 
 \label{fig:dexspin}
\end{figure}

Implementation of D-TDDFT requires several modifications of the
standard AA implementation of Casida's equation. First an algorithm to
decide which 2h2p excitations have to be included is needed. At the
present time, the user specifies the number of such excitations. These
are then automatically selected as the N lowest-energy 2h2p IPA
states. Since we are using a truncated 2h2p space, the algorithm makes
sure that all the spin partners are present, in order to have pure
spin states. The basic idea is illustrated in Fig.~\ref{fig:dexspin}.
Both 2h2p excitations are needed in order to construct the usual
singlet and triplet combinations. A similar algorithm should be
implemented for including all space double excitations which involve
degenerate irreducible representations, but this is not implemented in
the present version of the code.

These IPA 2h2p excitations are then added to the initial guess for the
Davidson diagonalizer. We recognize that a perturbative pre-screening
of the 2h2p space would be a more effective way for selecting the
excitations, but this more elaborate implementation is beyond the
scope of the present study.

We need new integrals to implement the HF exchange terms appearing in
the many-body theory blocks. The construction of these blocks require
extra hole-hole and particle-particle three-center integrals apart
from the usual hole-particle integrals already needed in TDDFT. We
then construct the additional matrix elements using the
resolution-of-the-identity (RI) formula
\begin{eqnarray}
  (pq|f|rs) &&= \\
  &&\sum_{IJKL}{(pq|g_I)S_{IJ}^{-1}(g_J|f|g_K)S_{KL}^{-1}(g_L|rs)} \nonumber \,
  ,
\end{eqnarray}
where $g_I$ are the usual {\sc deMon2k} notation for the density
fitting functions and $S_{IJ}$ is the auxiliary function overlap
matrix defined by $S_{IJ}=(g_I|g_J)$, in which the Coulomb repulsion
operator is used as metric.

Solving Eq.(\ref{eq:lowdin}) means solving a non-linear set of
equations. This is less efficient than solving linear equations.  In
Ref.~\cite{HC10} it was shown that Eq.~(\ref{eq:lowdin}) comes from
applying the L\"owdin-Feshbach partitioning technique to
\begin{equation}
 \label{eq:explow} \left[ \begin{array}{cc} {\bf A}_{11} & {\bf
 A}_{12} \\ {\bf A}_{21} & {\bf
 A}_{22} \\ \end{array} \right] \left[ \begin{array}{c} {\bf X}_1 \\
 {\bf X}_2 \end{array} \right] = \omega \left[ \begin{array}{c} {\bf
 X}_1 \\ {\bf X}_2 \end{array}\right] \, ,
\end{equation}
where $X_1$ and $X_2$ are now the single and double excitation
components of the vectors.  The solution of this equation is easier
and does not require a self-consistent approach, albeit at the cost of
requiring more physical memory, since then the Krylov space vectors
have the dimension of the single and the double excitation space.

Calculation of oscillator strengths has also to be modified when
D-TDDFT is implemented. In a mixed many-body theory and TDDFT
calculation, there is an extra term in the ground-state KS
wavefunction \cite{HC10}
\begin{equation}
  \label{eq:ksbrillouin} |0\rangle = \left( 1
  + \sum_{ia}{\frac{(i|{\hat M}_{xc} |a)}{\epsilon_i-\epsilon_a}{\hat
  a}_a^\dagger{\hat a_i}} \right) | KS \rangle \,
\end{equation}
where $|KS\rangle$ is the reference KS wavefunction. This equation
represents a ``Brillouin condition'' to the Kohn-Sham Hamiltonian.
The evaluation of transition dipole moments in {\sc deMon2k} was
modified to include the contributions from 2h2p poles,
\begin{eqnarray}
&& ({\bf r}|{\hat a}^\dagger_a{\hat a}_i{\hat a}^\dagger_b{\hat a}_j)
  = \nonumber \\ && X_{aibj} \left( \frac{(i|{\hat
  M}_{xc}|a)}{\epsilon_i-\epsilon_a}(j|{\bf r}|b) + \frac{(j|{\hat
  M}_{xc}|b)}{\epsilon_j-\epsilon_b}(i|{\bf r}|a) \right.\nonumber \\
  &-&\left. \frac{(i|{\hat M}_{xc}|b)}{\epsilon_i-\epsilon_b}(j|{\bf
  r}|a) -\frac{(j|{\hat M}_{xc}|a)}{\epsilon_j-\epsilon_a}(i|{\bf
  r}|b) \right) \, ,
\end{eqnarray}
where $X_{aibj}$ is an element of the eigenvector ${\bf X}_2$, the
double excitation part of the eigenvector of Eq.~(\ref{eq:explow}).


%% file: details.tex
Geometries for the set of 28 organic chromophores were taken from
Ref.~\cite{SSS+08a}.  These were optimized at the MP2/6-31G* level,
forcing the highest point group symmetry in each case. The orbital
basis set is Ahlrich's TZVP basis \cite{SHA92}.  As pointed out in
Ref.~\cite{SSS+08a}, this basis set has not enough diffuse functions
to converge all Rydberg states. We keep the same basis set for the
sake of comparison with the benchmark results. Basis-set errors are
expected for states with a strong valence-Rydberg character or states
above 7~eV, which are in general of Rydberg nature.

Comparison of the D-TDDFT is performed against the best estimates
proposed in Ref.~\cite{SSS+08a}. In each particular case the best
estimates might correspond to a different level of theory. If
available in the literature, these are taken as highly
correlated \textit{ab initio} calculations using large basis sets.  In
the absence, they are taken as the coupled cluster CC3/TZVP
calculation if the weight of the 1h1p space is more of than 95\%, and
CASPT2/TZVP in the other cases.

All calculations were performed with a development version of {\sc
deMon2k} (unless otherwise stated) \cite{deMon2k}. Calculations were
carried out with the fixed fine option for the grid and the GEN-A3*
density fitting auxiliary basis. The convergence criteria for the SCF
was set to $10^{-8}$.

To set up the notation used in the rest of the article, excited state
calculations are denoted by TD/SCF, where SCF is the functional used
for the SCF calculation and TD is the choice of post-SCF excited-state
method. Additionally, the D-TD/SCF($n$) and x-D-TD/SCF($n$) will refer
to the dressed and extended dressed TD/SCF method using $n$ 2h2p
states.  Thus TDA D-ALDA/AC-LDA(10) denotes a asymptotically-corrected
LDA for the DFT calculation followed by a LR-TDDFT calculation with
the dressed xc-kernel kernel and the Tamm-Dancoff approximation. The
D-TDDFT kernel has the adiabatic LDA xc-kernel for the ${\bf A}_{11}$
block and the ${\bf A}_{22}$ block is approximated as KS orbital
energy differences.

In this work, all calculations are done in using the TDA and a AC-LDA
wavefunction. For the sake of readability, we might omit writing them
when our main focus is on the discussion of the different variants of
the post-SCF part.

Calculations on our test-set show few differences between ALDA/LDA and
ALDA/AC-LDA. The singlet and triplet excitation energies and the
oscillator strengths are shown in Table~\ref{tab:results} of
Appendix~\ref{app:tables}. The average absolute error is 0.16~eV with
a standard deviation of 0.19~eV. The maximum difference is
0.91~eV. The states with larger differences justify the use of
asymptotic correction.  However, the absolute error and the standard
deviation are small. We attribute this to the restricted nature of the
basis set used in the present study.


%% file: results.tex
In this section we discuss the results obtained with the different
variants of D-TDDFT.  In particular, we compare the quality of D-TDDFT
singlet excitation energies against benchmark results for 28 organic
chromophores.  These chromophores can be classified in four groups
according to the chemical nature of their bond: (i) unsaturated
aliphatic hydrocarbons, containing only carbon-carbon double bonds;
(ii) aromatic hydrocarbons and heterocycles, including molecules with
conjugated aromatic double bonds; (iii) aldehydes, ketones and amides
with the characteristic oxygen-carbon double bonds; (iv) nucleobases
which have a mixture of the bonds found in the three previous groups.

These molecules have two types of low-lying excited states: Rydberg
(i.e., diffuse states) and valence states. The latter states are
traditionally described using the familiar H\"uckel model. The
low-lying valence transitions involve mainly $\pi$ orbitals, i.e. the
molecular orbitals (MO) formed as combinations of $p_z$ atomic
orbitals. The $\pi$ orbitals are delocalized over the whole
structure. Electrons in these orbitals are easily promoted to an
excited state, since they are not involved in the skeletal
$\sigma$-bonding. The most characteristic transitions in these systems
are represented by 1h1p $\pi\rightarrow\pi^*$ excitations. Molecules
containing atoms with lone-pair electrons can also have
$n\rightarrow\pi^*$ transitions, in which $n$ indicates the MO with a
localized pair of electrons on a heteroatom. In a few cases, we can
also have $\sigma\rightarrow\pi^*$ single excitations, although these
are exotic in the low-lying valence region.

\begin{figure}
  \caption{Schematic representation of the interaction between the
   1h1p and the 2h2p spaces. The relaxation energy $\Delta$ is
   proportional to the size of the coupling and inversely proportional
   to the energy difference between the two
   spaces.}  \includegraphics[scale=1.5]{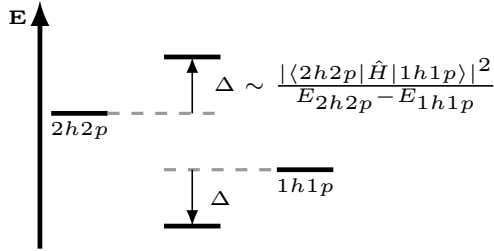} 
  \label{fig:correlation}
\end{figure}

The role of 2h2p (in general $n$h$n$p) poles is to add correlation
effects to the single excitation picture. For the sake of discussion,
it is important to classify (loosely) the correlation included by 2h2p
states as static and dynamic.  Static correlation is introduced by
those double excitations having a contribution similar to the single
excitations for a given state. This requires that the 1h1p excitations
and the 2h2p excitations are energetically near and have a strong
coupling between the two (Fig.~\ref{fig:correlation}.)  We will refer
to such states as multireference states. Dynamical correlation is a
subtler effect.  Its description requires a much larger number of
double excitations, in order to represent the cooperative movement of
electrons in the excited state.

For the low-lying multireference states found in the molecules of our
set, a few double excitations are required for an adequate first
approximation.  Organic chromophores of the group (i) and (ii) have a
characteristic low-lying multireference valence state (commonly called
the $L_b$ state in the literature) of the same symmetry as the
ground-state. The $L_b$ state is well known for having important
contributions from double excitations of the type
$(\pi_\alpha,\pi_\beta) \rightarrow (\pi^*_\alpha,\pi^*_\beta)$,
thereby allowing mixing with the ground state. Some contributions of
double excitations from $\sigma$ orbitals might also be important to
describe relaxation effects of the orbitals in the excited state that
cannot be accounted by the self-consistent field orbitals \cite{A10}.

\begin{figure}
  \caption{Dependence of the 1h1p triplet (solid line) and singlet
           (dashed line) excitation energies of one excitation of
           ethene with increasing number of double
           excitations. Calculations are done with D-ALDA/AC-LDA.
           Excitation energies are in
           eV.}  \includegraphics[scale=0.9]{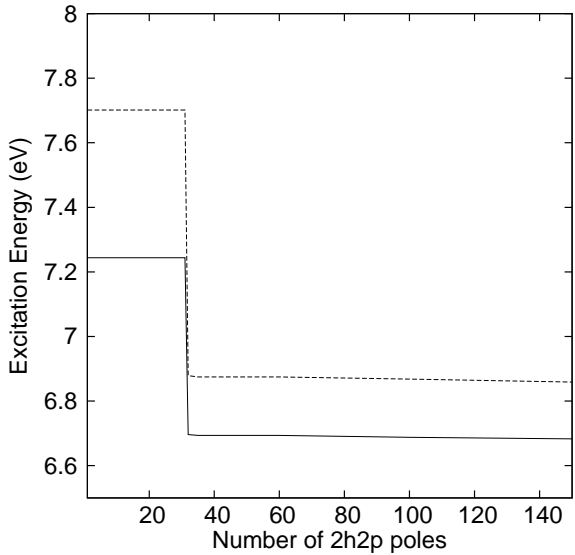} 
  \label{fig:double}
\end{figure}

The different effects of the 2h2p excitations that include dynamic and
static correlation are clearly seen in the changes of the 1h1p
adiabatic energies when we increase the number of double
excitations. As an example, we take two states of ethene, one triplet
and singlet 1h1p excitations, for which we systematically include a
larger number of 2h2p states. The results for the D-ALDA/AC-LDA
approach are shown in Fig.~\ref{fig:double}. We plot the adiabatic
1h1p states for which we include one 2h2p excitation at a time until
35, after which the steps are taken adding ten 2h2p states at a
time. When a few 2h2p states are added, we observe that the excitation
energy remains constant. This is probably due to the high symmetry of
the molecule, which 2h2p states are not mixed with 1h1p states by
symmetry selection rules.  It is only when we add 32 double
excitations when we see a sudden change of the excitation energy of
both triplet and singlet states. This indicates that we have included
in our space the necessary 2h2p poles to describe the static
correlation of that particular state.  Static correlation has a major
effect in decreasing the excitation energy with a few number of 2h2p
excitations. In this specific case, the triplet excitation energy
decreases by 0.54~eV while the singlet excitation energy decreases by
0.82~eV. In this case, all static 2h2p poles are added, and a larger
number of these poles does not lead to further sudden changes. The
excitations are almost a flat line, with a slowly varying slope.  This
is the effect of the dynamic correlation, which includes extra
correlation effects but which does not suddenly vary the excitation
energy.

A-TDDFT includes some correlation effects in the 1h1p states, both of
static and dynamic origin. However, it misses completely the states of
main 2h2p character. These states are explicitly included by the
D-TDDFT kernel. Additionally, D-TDDFT includes extra correlation
effects into the A-TDDFT 1h1p states through the coupling of 1h1p
states with the 2h2p states.  This can lead to double counting of
correlation, i.e., the correlation already included by A-TDDFT can be
reintroduced by the coupling with the 2h2p states, leading to an
underestimation of the excited state. In order to avoid double
counting of correlation, it is of paramount importance to have a deep
understanding of which correlation effects are included in each of the
blocks that are used to construct the D-TDDFT xc-kernel. Therefore, we
have compared the different D-TDDFT kernels with a reference method of
the same level of theory, but from which the results are well
understood.  This is provided by some variations of the \textit{ab
initio} method CISD, since the mathematical form of the equations is
equivalent to the TDA approximation of D-TDDFT. Standard CISD has
coupling with the ground state, which we have not included in D-TDDFT.
Therefore, we have made some variations on the standard CISD
(Sec.~\ref{sec:dressed}.) We call these variations D-CIS and x-D-CIS,
according to the definition of the ${\bf A}_{22}$ block. In both
methods, the 1h1p block ${\bf A}_{11}$ is given by the CIS
expressions, which does not include any correlation effect (recall
that in response theory, correlation also appears in the
singles-singles coupling block.)  The correlation effects in D-CIS and
x-D-CIS are included only through the coupling between 1h1p and 2h2p
states. This will provide us with a good reference for rationalizing
the results of A-TDDFT versus D-TDDFT.

Our implementation of CIS and (x-)D-CIS is done in {\sc deMon2k}.
Therefore, all CI calculations actually refer to RI-CI and are based
on a DFT wavefunction. We have calculated the absolute error between
HF-based CIS excitation energies (performed with {\sc
Gaussian} \cite{g03}) and CIS/AC-LDA excitation energies for the
molecules in the test set. We have found little differences
(Appendix~\ref{app:tables}), giving an average absolute error is
0.18~eV with a standard deviation of 0.13~eV and a maximum absolute
difference of 0.54~eV. It is interesting to note that almost all
CIS/AC-LDA excitations are slightly below the corresponding HF-based
CIS results.

\begin{figure*}[!]
  \caption{Correlation graphs of singlet excitation energies for
           different flavors of D-CIS and D-TDDFT with respect to best
           estimates. Excitation energies are given in
           eV.}  \begin{tabular}{cc} (a) CIS/AC-LDA & (b) TDA
           ALDA/AC-LDA \\
\includegraphics[scale=0.7]{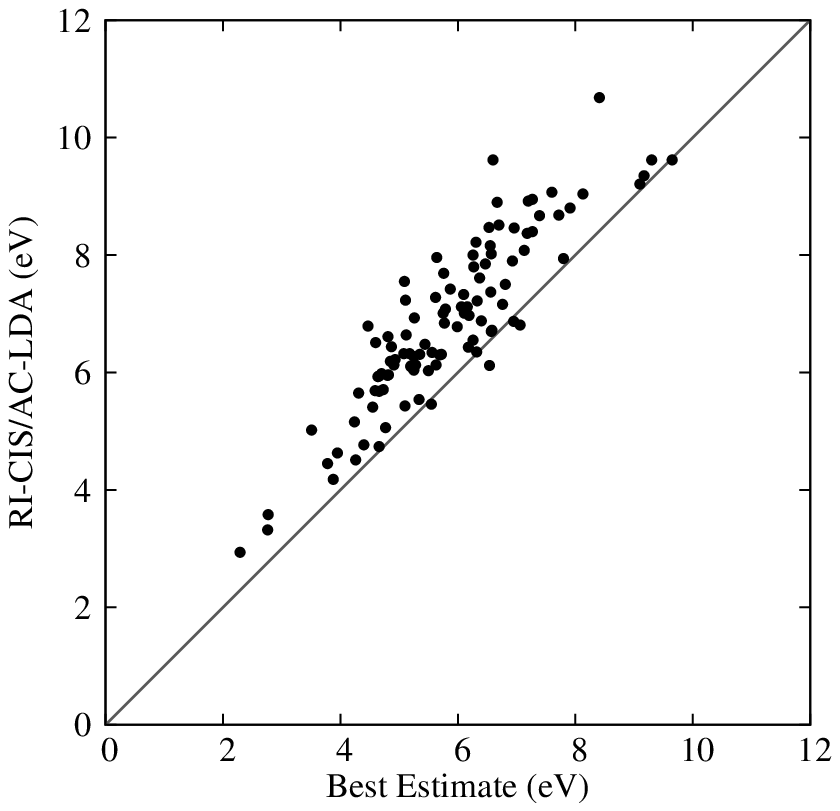} & \includegraphics[scale=0.7]{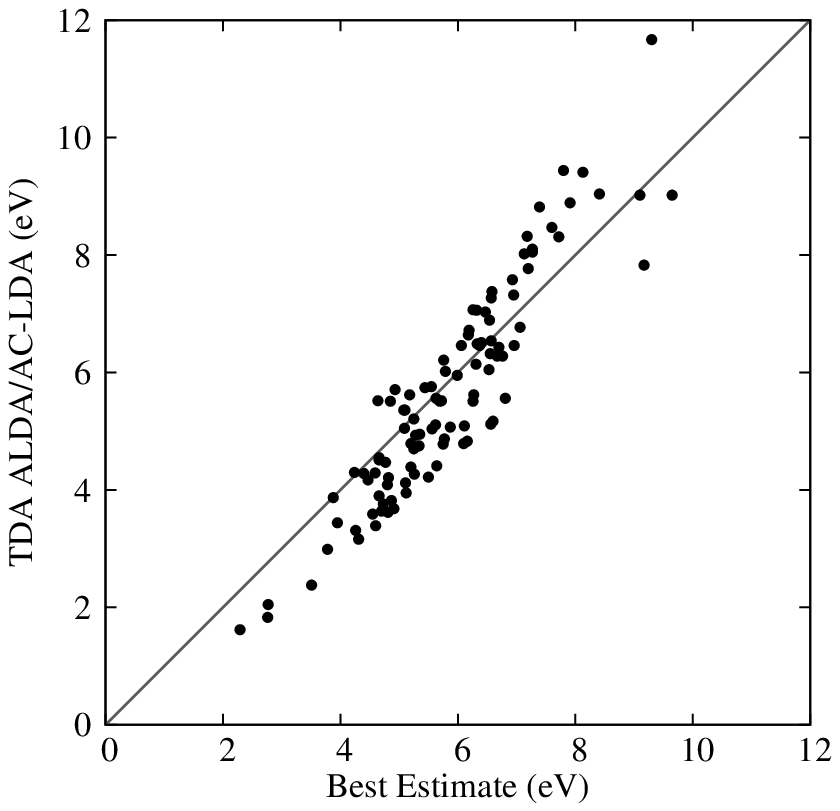} \\
(c) D-CIS/AC-LDA(10) & (d) TDA D-ALDA/AC-LDA(10) \\
\includegraphics[scale=0.7]{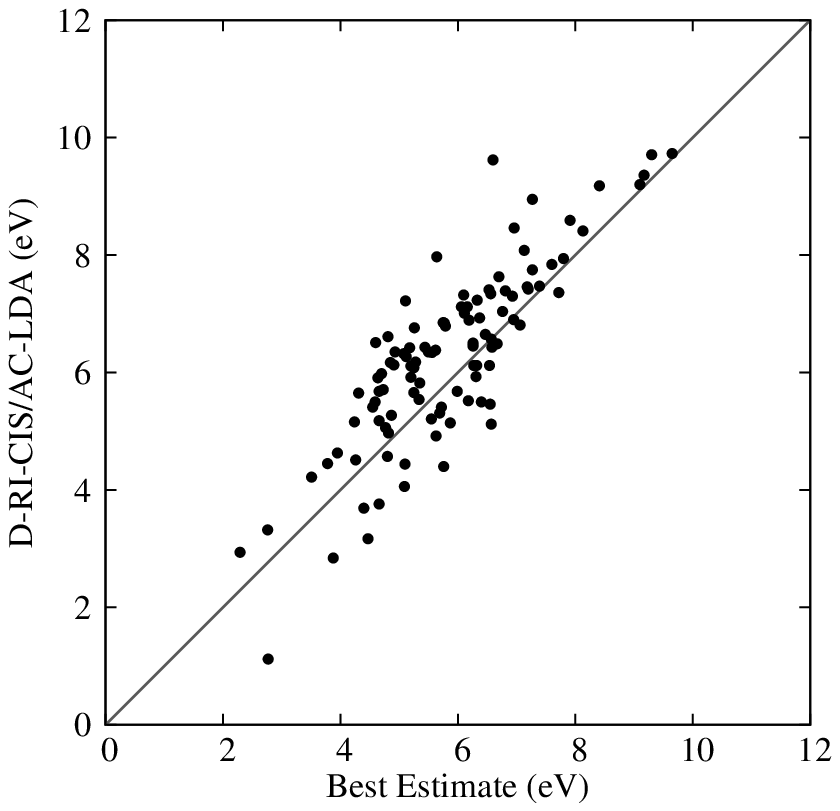} & \includegraphics[scale=0.7]{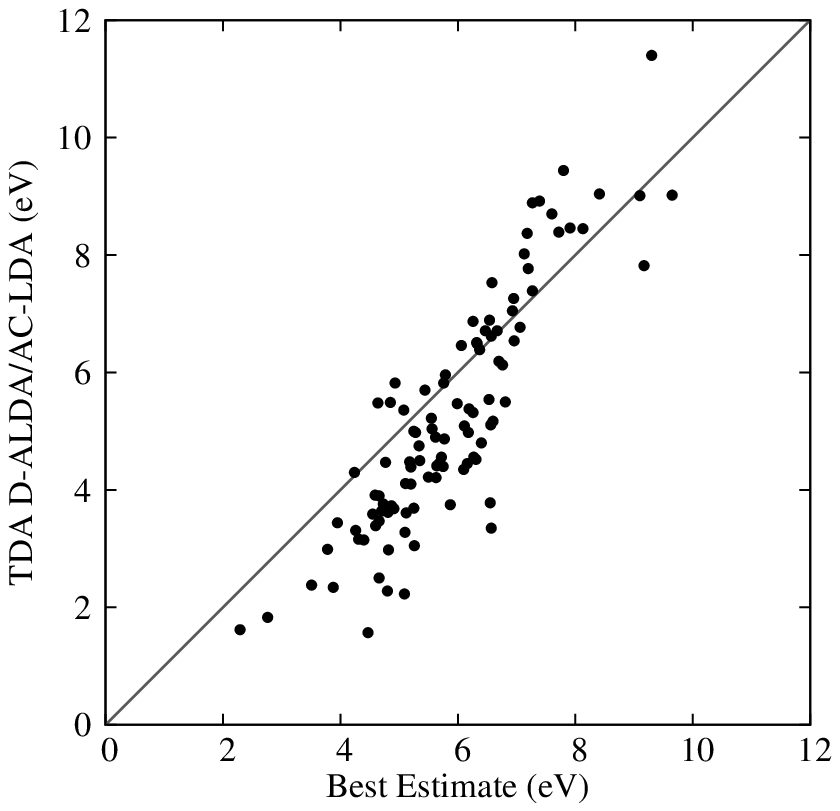} \\
(e) x-D-CIS/AC-LDA(10) & (f) TDA x-D-ALDA/AC-LDA(10) \\
\includegraphics[scale=0.7]{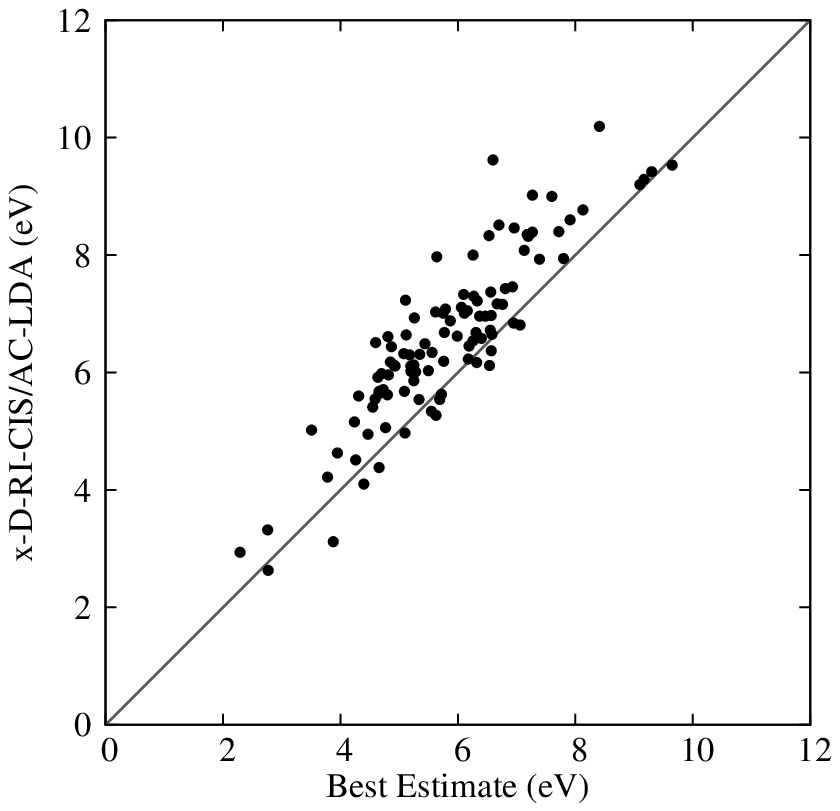} & \includegraphics[scale=0.7]{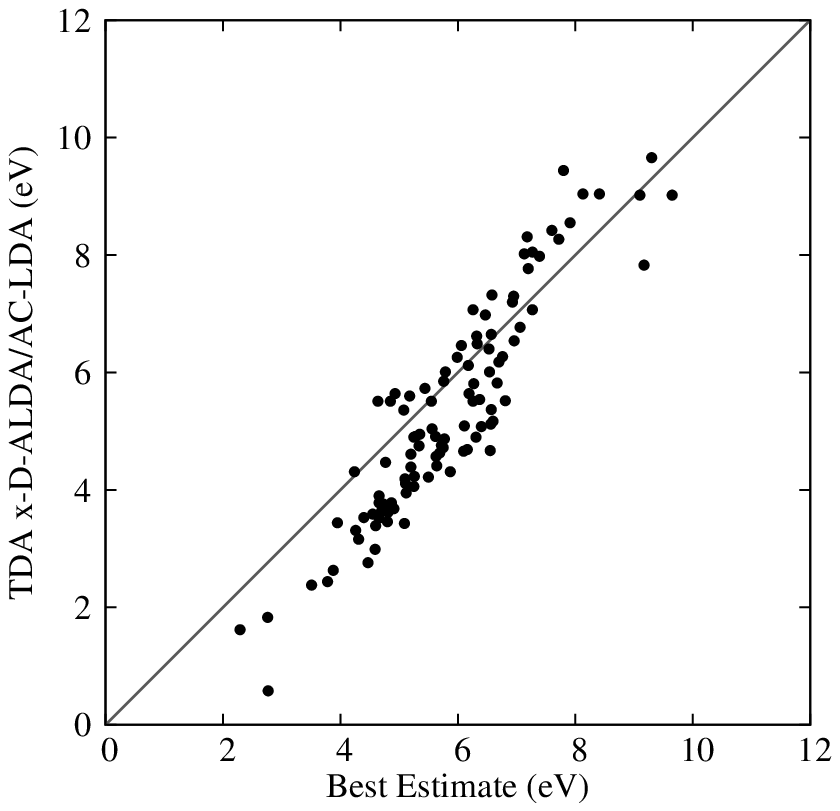} \\
  \end{tabular} 
  \clearpage
  \label{fig:singlets}
\end{figure*}

We now discuss the results for singlet excitation energies of A-TDDFT
and D-TDDFT. Since the number of states is large, we will discuss only
general trends in terms of correlation graphs for each of the methods
used with respect to the benchmark values provided in
Refs.~\cite{SSS+08a,SSS+08b}.  Our discussion will mainly focus on
singlet excitation energies.  For the numerical values of triplets,
singlets, and oscillator strengths for each specific molecule, the
reader is referred to Table~\ref{tab:results} of
Appendix~\ref{app:tables}.

We first discuss the results of the adiabatic theories (i.e.,
$\omega$-independent) CIS/AC-LDA and TDA ALDA/AC-LDA, shown in graphs
(a) and (b) of Fig.~\ref{fig:singlets} respectively. None of these
theories includes 2h2p states, although ALDA includes some correlation
effects in the 1h1p states through the xc-kernel.  We see that CIS
overestimates all excitation energies with respect to the best
estimates. This is consistent with the fact that CIS does not include
any correlation effects. The mean absolute error is 1.04~eV with a
standard deviation of 0.63~eV.  The maximum error is 3.02~eV. A better
performance of ALDA is observed. We see that ALDA underestimates most
of the excitation energies, especially in the low-energy region. A
similar conclusion was drawn by Silva-Junior \textit{et
al.} \cite{SSS+08b}, who applied the pure BP86 xc-kernel to the
molecules of the same test set. Nonetheless, the overall performance
of ALDA is clearly superior over CIS, giving an average absolute error
of 0.67~eV with a standard deviation of 0.44~eV.  The maximum absolute
error of is 2.37~eV.

When we include explicit double excitations in CIS and A-TDDFT, we
include correlation effects to the 1h1p picture and the excitation
energies decrease. We have truncated the number of 2h2p states to 10
double excitations, in order to avoid the double counting of
correlation in the D-TDDFT methods and in order to keep the
calculations tractable. However, we realize that with our primitive
implementation, the use of only 10 2h2p states may not include all
static correlation necessary to correct all the states, especially for
higher-energy 1h1p states.

As we have shown in Sec.~\ref{sec:dressed}, there is more than one way
to include the 2h2p effects. We first consider the D-CIS/AC-LDA(10)
and TDA D-ALDA/AC-LDA(10) variants, shown in graphs (c) and (d) of
Fig.~\ref{fig:singlets}, in which we approximate the double-double
block by a diagonal zeroth-order KS orbital energy difference. In both
cases, we observe that the results get worse with respect to those of
CIS or ALDA. This degradation is especially important for D-ALDA(10)
and might be interpreted as due to double counting of
correlation. Already, ALDA underestimates the excitation energies of
most states. With the introduction of double excitations, we introduce
extra correlation effects, which underestimates even more the
excitations. In some cases, like \textit{o}-benzoquinone
(Appendix~\ref{app:tables}), some excitation energies falls below the
reference ground-state, possibly indicating the appearance of an
instability. The average absolute error of the D-ALDA(10) is 1.03~eV
with a standard deviation of 0.73~eV and a maximum error of 3.51~eV,
decreasing the description of 1h1p states with respect to ALDA or CIS.
As to D-CIS(10), the results are slightly better. The average absolute
error is 0.78~eV with a standard deviation of 0.54~eV and a maximum
error of 3.02~eV, improving over the CIS results.  However, some
singlet excitation energies are smaller than the corresponding triplet
excitation energies and some state energies are now largely
underestimated.  This also indicates an overestimation of correlation
effects, though it might be partially due to the missing $A_{02}$
block.

A better estimate of the 2h2p correlation effects is given when the
${\bf A}_{22}$ block is approximated with first-order correction to
the HF orbital energy differences. This type of calculation is what we
call x-D-CIS/AC-ALDA(10) and x-D-TDDFT/AC-ALDA(10), the results of
which are shown respectively in graphs (e) and (f) of
Fig.~\ref{fig:singlets}. In both cases we observe an improvement of
the excitation energies. The x-D-CIS provides a more consistent and
systematic estimation of correlation effects, and most of the
excitations are still an upper limit to the best estimate result.
However, the mean absolute error is still high, with an average
absolute error of 0.84~eV and a standard deviation of 0.58~eV and a
maximum error of 3.02~eV.  The x-D-TDDFT results slightly improve over
x-D-CIS, giving a mean absolute error of 0.83~eV with a standard
deviation of 0.46~eV and a maximum error of 2.19~eV. The superiority
of x-D-TDDFT is explained by the fact that TDDFT includes some
correlation effects in the 1h1p block.  However, x-D-TDDFT still gives
in overall larger errors than A-TDDFT. This might be again a problem
of double-counting of correlation. Since A-TDDFT with the ALDA
xc-kernel underestimates most excitation energies, the application of
x-D-TDDFT leads to a further underestimation. In any case, D-TDDFT
works better when 2h2p states are given by the first-order correction
to the HF orbital energy difference.

From the schematic representation of the interaction between 1h1p
states and 2h2p states (Fig.~\ref{fig:correlation}), we can
rationalize why we observe overestimation of correlation when the
${\bf A}_{22}$ block approximated as an LDA orbital energy
difference. The 2h2p states as given by the LDA fall too close
together and too close to the 1h1p states (i.e., a too large value of
$\Delta$). The results show large correlation effects in the 1h1p
states, indicating an overestimation of static correlation effects.
The first-order correction to the KS orbital energy difference give a
better estimate of correlation effects. The reversed effect was
observed in the context of HF-based response theory. In SOPPA
calculations, the 2h2p states are approximated as simple HF orbital
energy differences, which are placed far too high, therefore
underestimating correlation. In HF-based response, it was also seen
that the results are improved when adding the first-order correction
to the HF orbital energy differences.

\begin{figure}
  \caption{A-TDDFT and x-D-TDDFT correlation graphs for singlet 
           excitation energies using the hybrid xc-kernel of 
           Eq.~(\ref{eq:hybrid}), in which $c_0=0.2$.} 
   \begin{tabular}{l}
     (a) TDA HYBRID/AC-ALDA \\
     \includegraphics[scale=0.7]{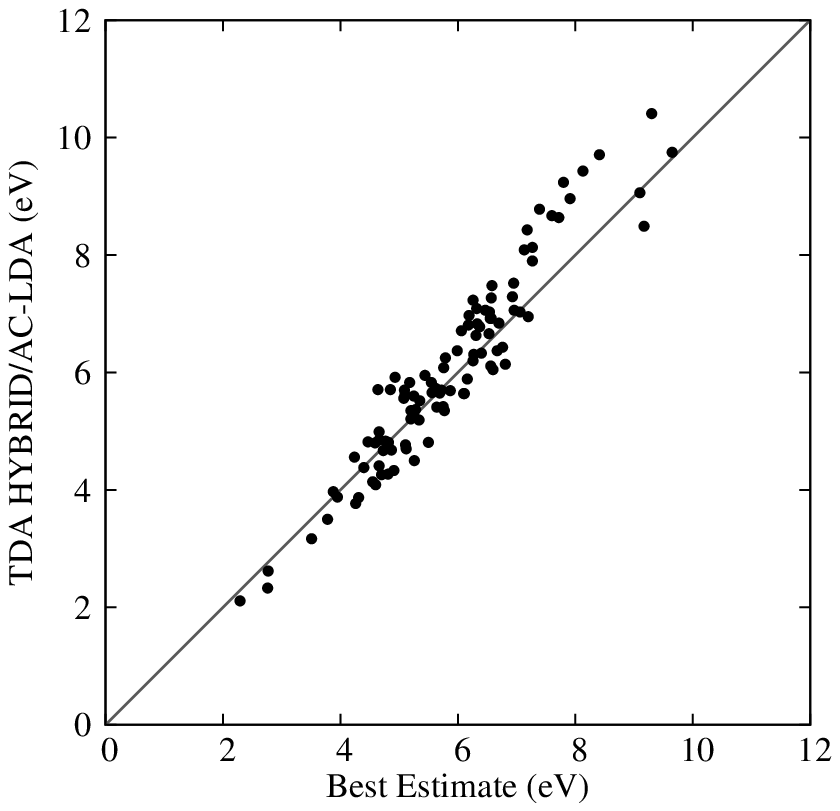} \\
     (b) TDA x-D-HYBRID/AC-ALDA \\
     \includegraphics[scale=0.7]{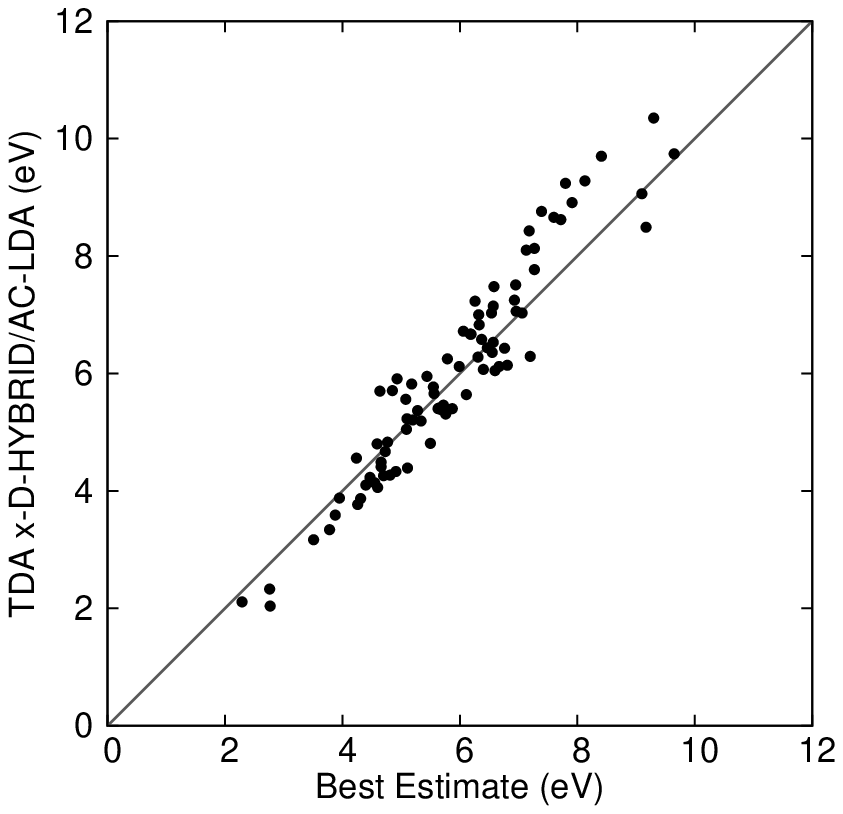} \\
   \end{tabular}
  \label{fig:hybrid}
\end{figure}

Up to this point, we have seen that D-TDDFT works best when 2h2p
states are given by the first-order correction to the HF orbital
energy differences.  However, we have also seen that the LDA xc-kernel
underestimates the 1h1p states, so that we degrade the quality of the
A-TDDFT states when we apply any of the D-TDDFT schemes.  A better
estimate for the 1h1p states is given by an adiabatic hybrid
calculation.  In Fig.~\ref{fig:hybrid}~(a) we show the calculation of
our implementation of the hybrid xc-kernel based upon a LDA
wavefunction. In this hybrid we use 20\% HF exchange. The results show
an improvement over all our previous calculations. The average
absolute error of 0.43~eV with respect to the best estimates and a
standard deviation of 0.34~eV. The maximum error is
1.44~eV. Figure~\ref{fig:hybrid}~(b) shows the x-D-HYBRID(10)
calculation. The mean error and the standard deviation are very
similar to what the adiabatic hybrid calculation gives. The average
absolute error with respect to the best estimate is 0.45~eV, and the
standard deviation is 0.33~eV with a maximum error of 1.44~eV. This is
a very important result, since we have been able to include the
missing 2h2p states without decreasing the quality of 1h1p states.

\begin{table}
  \caption{Summary of the mean absolute errors, standard deviation and
           maximum error of each method. All quantities are in eV.}
  \begin{tabular}{cccc}
    \hline
      Method          & Mean error & Std. dev. & Max. error \\ 
    \hline
      ALDA            & 0.67       & 0.44      & 2.37 \\ 
      D-ALDA(10)      & 1.03       & 0.73      & 3.51 \\ 
     x-D-ALDA(10)     & 0.83       & 0.46      & 2.19 \\ 
      CIS             & 1.04       & 0.63      & 3.02 \\ 
      D-CIS(10)       & 0.78       & 0.54      & 3.02 \\ 
      x-D-CIS(10)     & 0.84       & 0.58      & 3.02 \\ 
      HYBRID          & 0.43       & 0.34      & 1.44 \\ 
      x-D-HYBRID(10)  & 0.45       & 0.33      & 1.44 \\ 
    \hline
  \end{tabular}
  \label{tab:summaryerror}
\end{table}

In Table~\ref{tab:summaryerror} we summarize the mean absolute errors,
standard deviations and maximum errors for all the methods. The best
results are given by the hybrid A-TDDFT calculation, closely followed
by the x-D-TDDFT based also on the hybrid. We can therefore state that
the best D-TDDFT kernel can be constructed from a hybrid xc-kernel in
the ${\bf A}_{11}$ block and the first-order correction to the HF
orbital energy differences for ${\bf A}_{22}$.

The results given by the different D-TDDFT kernels show a close
relation between the ${\bf A}_{11}$ and ${\bf A}_{22}$ blocks.  Our
results show that the singles-singles block is better given by a
hybrid xc-kernel and the doubles-doubles block is better approximated
by the first-order correction to the HF orbital energy difference. By
simple perturbative arguments, we have rationalized that the ${\bf
A}_{22}$ block as given by the first-order approximation accounts
better for static correlation effects. Less clear explanations can be
given to understand why a hybrid xc-kernel gives the best
approximation for the ${\bf A}_{11}$ block, although it seems
necessary for the construction of a consistent kernel.
\begin{figure}
  \caption{Effect on excited states with more than 10\% of 2h2p
    character of mixing HF exchange in TDDFT. CASPT2 results from
    Ref.~\cite{SSS+08a} are taken as the benchmark. BHLYP results are
    taken from Ref.~\cite{SSS+08b}.}  \begin{tabular}{c} (a) Single
    excitations with CIS and
    A-TDDFT \\ \includegraphics{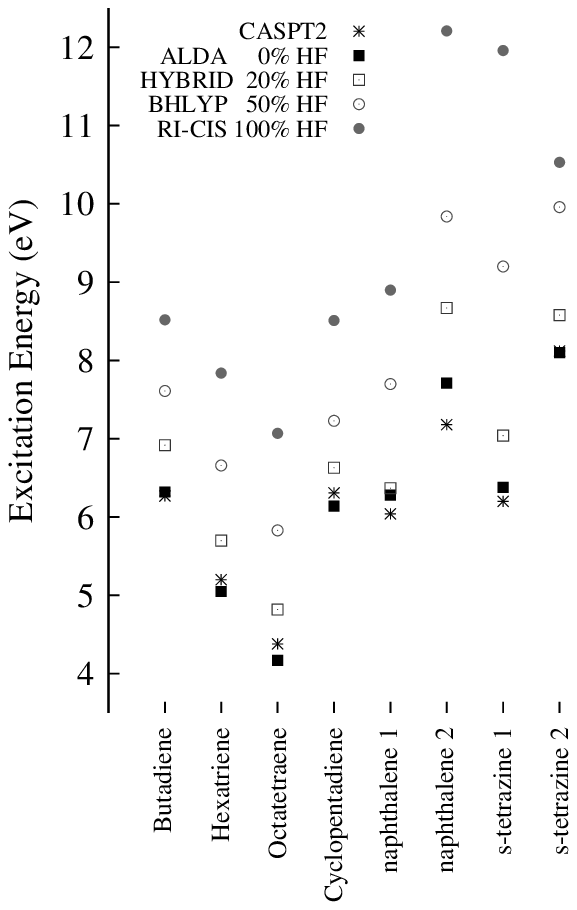} \\ (b) Single
    excitations with D-CIS and
    D-TDDFT \\ \includegraphics{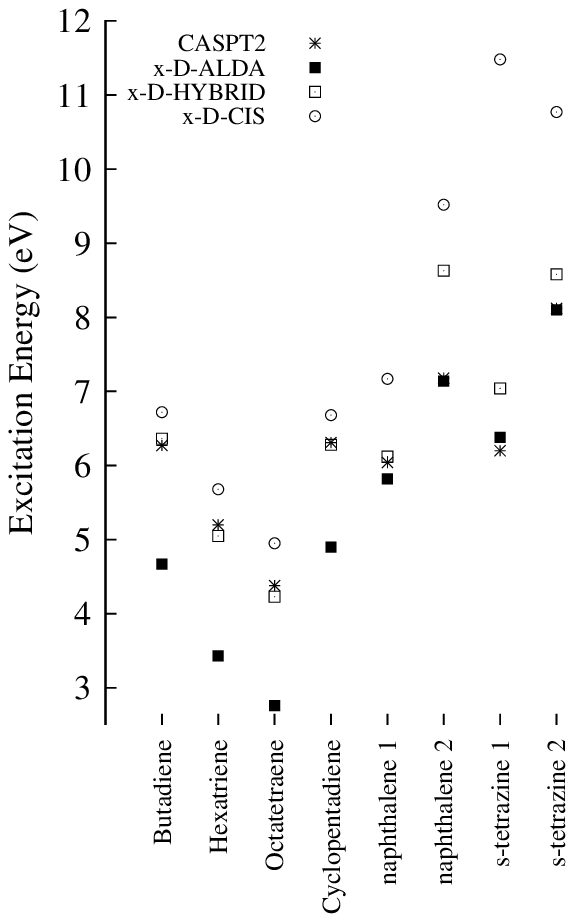} \end{tabular} 
  \label{fig:lb}
\end{figure}

The main interest of using a D-TDDFT kernel is to obtain the pure 2h2p
states, which are not present in A-TDDFT and to better describe the
1h1p states of strong multireference character. We now take a closer
look at the latter states in our test set. In particular, we will
compare against the benchmarks those 1h1p states that have a 2h2p
contribution larger than 10\% (this percentage is determined by the
CCSD calculation of Ref.~\cite{SSS+08a}.)  The molecules containing
such states are the four polyenes of the set, together with
cyclopentadiene, naphthalene and \textit{s}-triazine. From this
sub-set, the polyenes are undoubtedly the ones which have been the
most extensively discussed. Some debate persists as to whether A-TDDFT
is able to represent a low-lying localized valence state which have a
strong 2h2p contribution of the transition promoting two electrons
from the highest- to the lowest-occupied molecular orbital. It was
first shown by Hsu \textit{et al.} that A-TDDFT with pure functionals
gives the best answer for such states \cite{HHH01}, catching both the
correct energetics and the localized nature of the state. Starcke
\textit{et al.} recognize this to be a fortuitous cancellation of 
errors \cite{SWS+06}.

In the top graph of Fig.~\ref{fig:lb} we show the the behavior of CIS
(100\% HF exchange) and A-TDDFT with different hybrids: ALDA with 0\%
HF exchange, ALDA with 20\% HF exchange and BHLYP which has 50\% HF
exchange. In this comparison, we take the CASPT2 results (stars) as
the benchmark result, since the best estimates were not provided for
all the studied states \cite{SSS+08a}. As seen in the graph, CIS
(filled circles) seriously overestimate the excitation energies,
consistent with the fact that it does not include any correlation
effect. A-TDDFT with pure functionals give the best answer for
doubly-excited states, very close to the CASPT2 result. This confirms
the observation of Hsu \textit{et al.} \cite{HHH01} Hybrid
functionals, though giving the best overall answer, do not perform as
good for these states. Additionally, the more HF exchange is mixed in
the xc-kernel, the worse the result is.  A different situation appears
when we include explicitly 2h2p states. In Fig.~\ref{fig:lb}~(b), we
show the results of x-D-CIS and x-D-TDDFT. Now, the x-D-ALDA(10)
underestimates the multireference excitation energies, due to
overcounting of correlation effects. The best answer is now given by
x-D-HYBRID(10) with 20\% HF exchange. The x-D-CIS stays always
higher. One can notice that the three last excitations (naphthalene 2
and \textit{s}-triazine 1 and 2) are best described by the
x-D-ALDA(10). This can be simply due to the fact that we missed the
important double excitation to represent these states, since we
restrict our calculation to 10 2h2p states and we add them in strict
energetic order with no pre-screening.


%% file: conclude.tex

D-TDDFT was introduced by Maitra \textit{et al.} to explicitly include
2h2p states in TDDFT.  The original work was \textit{ad hoc}, leaving
much room for variations on the original concept.  A limited number of
applications by Maitra and coworkers \cite{MZC+04,CZM+04} as well as
by Mazur \textit{et al.} \cite{MW09,MMR+10} showed promising results
for D-TDDFT, but could hardly be considered definitive because (i) of
the limited number of molecules and excitations treated and (ii)
because the importance of the details of the specific implementations
of D-TDDFT were not adequately explained. The present article has gone
far towards remedying these problems, and providing further support
for D-TDDFT.

We have implemented several variations of D-TDDFT and RI-CI in {\sc
deMon2k}, with the aim of characterizing the minimum necessary
ingredients for an effective implementation of D-TDDFT. We have seen
that DFT-based CIS gives very similar answers to HF-based CIS, showing that the
effects of exact (HF) exchange can indeed be added in a post-SCF calculation. We have
also found that although ALDA works better than CIS, it
underestimates most of the excitation energies. Therefore, when we
explicitly include 2h2p states through D-TDLDA, it leads to worse
results, due to the double counting of correlation. The x-D-ALDA
give least scatter of the results and hence a better
answer. Nevertheless, the lower errors are still given by ALDA.

With the results of ALDA, we have shown that it is important to have
a correct relative position of the 1h1p space and the 2h2p space in
order to have a consistent account of correlation. We have introduced
a hybrid TDDFT as a post-LDA calculation, and we have shown that the
results are superior to those of ALDA. We have determined that the
method giving the best answer for MR states is the combination of a
hybrid xc-kernels with the  2h2p double
excitations approximated with first-order corrections to the HF
orbital energy differences.

Our work has gone much farther than previous work in testing D-TDDFT and
in detailing the necessary ingredients to make it work well, We find a
hybrid approach to be essential. We recognize that our work could be
improved by a perturbative pre-selection procedure and consider this
work to be ample justification for a more elaborate implementation of
D-TDDFT. This work also constitutes a key step towards a full
implementation of the polarization propatagor model of the exact
$f_{xc}(\omega)$.


%% file: thanks.tex
M.\ H.\ would like to acknowledge a scholarship from the French
Ministry of Education.  Those of us at the {\em Universit\'e Joseph
Fourier} would like to thank Denis Charapoff, R\'egis Gras,
S\'ebastien Morin, and Marie-Louise Dheu-Andries for technical support
at the (DCM) and for technical support in the context of the {\em
Centre d'Exp\'erimentation du Calcul Intensif en Chimie} (CECIC)
computers used for some of the calculations reported here.  This work
has been carried out in the context of the French Rh\^one-Alpes {\em
R\'eseau th\'ematique de recherche avanc\'ee (RTRA): Nanosciences aux
limites de la nano\'electronique} and the Rh\^one-Alpes Associated
Node of the European Theoretical Spectroscopy Facility (ETSF).  AR.\
acknowledges funding by the Spanish MEC (FIS2007-65702-C02-01),
ACI-promciona project (ACI2009-1036), ``Grupos Consolidados UPV/EHU
del Gobierno Vasco'' (IT-319-07), the European Research Council
through the advance grant DYNamo (267374), and the European Community
through projects e-I3 ETSF (Contract No. 211956) and THEMA (228539).


%% file: soppa.tex
In this appendix, we summarize the main expressions for the
construction of the matrix elements of Eqs.~(\ref{eq:soppakernel}) and
(\ref{eq:explow}). For
a detailed derivation, the reader is referred to 
Ref.~\cite{HC10}, in which this equations were derived for
the construction of an exact \textit{ab initio} xc-kernel 
consistent to second-order in perturbation theory.

The explicit expression for the single-single block is given by
\begin{eqnarray}
 && [A_{11}]_{ai,bj} = \\
  &&\left[\epsilon_a\delta_{ab}+(a|{\hat M}_{xc}|b) -\sum_{l}{\frac{(a|{\hat M}_{xc}|l)(l|{\hat M}_{xc}|b)}{\epsilon_{l}-\epsilon_{a}}}\right. \nonumber \\
&-&\left. \frac{1}{2}\sum_{mld}{\frac{(ld||mb)(dl||ma)}{\epsilon_m + \epsilon_l - \epsilon_d - \epsilon_a}} \right]\delta_{ij} \nonumber \\
  &-&\left[\epsilon_i\delta_{ij}+(i|{\hat M}_{xc}|j) -\sum_{d}{\frac{(i|{\hat M}_{xc}|d)(d|{\hat M}_{xc}|j)}{\epsilon_{i}-\epsilon_{d}}}\right. \nonumber \\
&-&\left. \frac{1}{2}\sum_{lke}{\frac{(le||jd)(dl||ei)}{\epsilon_i + \epsilon_l - \epsilon_d - \epsilon_e}} \right]\delta_{ab} \nonumber \, , 
\end{eqnarray}
the single-double block is given by
\begin{eqnarray}
  [A_{12}]_{ck,aibj} &=&  \delta_{kj}(bc||ai) - \delta_{ki}(bc||aj) \\
                     &+&  \delta_{ac}(bi||kj) - \delta_{bc}(ai||kj) \nonumber \, ,
\end{eqnarray}
and the double-double block is given by
\begin{equation}
  [A_{22}]_{aibj,ckdl} = \left(\epsilon_b + \epsilon_a - \epsilon_i - \epsilon_j \right)\delta_{ac}\delta_{ik}\delta_{bd}\delta_{jl} \, . 
\end{equation}
there $(pq||rs) = (pq|rs) - (qs|rq)$, where 
\begin{equation}
  (pq|rs) = \int{d^3rd^3r' \psi^*_p({\bf r}) \psi_q({\bf r})\frac{1}{|{\bf r}-{\bf r}'|}\psi^*_r({\bf r}') \psi^*_s({\bf r}')} \, .
\end{equation}

The first-order double-double block is given by
\begin{eqnarray}
  &&[A_{22}]_{aibj,ckdl} =  \\
&&\left[\left(\epsilon_b\delta_{bd} + (b|{\hat M}_{xc}|d) \right)\delta_{ac}
                          +\left(\epsilon_a\delta_{ac} + (a|{\hat M}_{xc}|c) \right)\delta_{bd} 
                           \right]\delta_{ik}\delta_{jl} \nonumber \\
   &-& \left[ \left(\epsilon_i\delta_{ik} + (i|{\hat M}_{xc}|k) \right)\delta_{jl} 
        -\left(\epsilon_j\delta_{jl} + (d|{\hat M}_{xc}|l) \right)\delta_{ik}\right]\delta_{ac}\delta_{bd} \nonumber \\
&-&\delta_{ac}f(bd)
  -\delta_{bd}f(ac)
  +\delta_{ad}f(bc)
  +\delta_{bc}f(ad)\nonumber\\
&-&\delta_{ac}\delta_{bd}(kj||li)
-\delta_{jl}\delta_{ki}(ad||bc) \nonumber \, ,
\end{eqnarray}
with
\begin{eqnarray}
f(pq) &=& \delta_{ik}(lj||pq) + \delta_{jl}(ki||pq) \nonumber \\
      &-& \delta_{kj}(li||pq) - \delta_{il}(kj||pq) \, .
\end{eqnarray}
Integrals with double bar are defined as in Eq.~(\ref{eq:intdef}), in
which the kernel $f$ is defined by 
$f({\bf r}_1,{\bf r}_2)=(1-{\hat P}_{12})/|{\bf r}_1-{\bf r}_2|$, where $P_{12}$
is the permutation operator that permutes the coordinates of two electrons.

%% file: tab.tex
\begin{table*}
\caption{Singlet and Triplet excitation energies and oscillator strengths. All excitation energies are in eV. The CASPT2, Best Estimates (Best) and B3LYP calculations are taken from Refs.~\cite{SSS+08a}~and~\cite{SSS+08b}. The HF-based CIS calculations (CIS) are done with {\sc Gaussian03} \cite{g03}. The rest are done in {\sc deMon2k} \cite{deMon2k}.}
\begin{tabular}{ccccccccccccc}
\hline
Ethene&CASPT2&Best&B3LYP&CIS&RI-CIS&D-CIS&x-D-CIS&ALDA&DALDA&x-D-ALDA&hybrid&x-D-hybrid\\
\hline
$1^1B_{1u}$&7.98&7.8&7.7&8.15&7.94&7.94&7.94&9.44&9.44&9.44&9.24&9.24\\
f&0.36&&0.362&0.633&0.59&0.59&0.59&0.507&0.507&0.507&0.558&0.558\\
$1^3B_{1u}$&4.39&4.5&4.03&3.46&3.26&3.26&3.26&5.95&5.95&5.95&5.54&5.55\\
Butadiene&&&&&&&&&&&&\\
$2^1A_g$&6.27&6.55&6.82&8.52&8.16&5.46&6.72&6.32&3.78&4.67&6.92&6.36\\
$1^1B_u$&6.23&6.18&5.74&6.55&6.43&5.52&6.23&6.64&4.98&6.12&6.81&6.67\\
f&0.686&&0.672&1.214&1.31&0.885&1.22&0.922&0.47&0.726&1.07&1.02\\
$1^3A_g$&4.89&5.08&4.86&4.26&4.25&4.25&4.25&6.21&6.21&6.21&6.12&6.12\\
$1^3B_u$&3.2&3.2&2.76&2.48&2.12&1.94&2.06&4.08&3.31&3.81&3.9&3.83\\
Hexatriene&&&&&&&&&&&&\\
$2^1A_g$&5.2&5.09&5.69&7.84&7.55&4.06&5.68&5.05&2.23&3.43&5.7&5.05\\
$1^1B_u$&5.01&5.1&4.69&5.56&5.43&4.44&4.97&5.36&3.28&4.19&5.62&5.23\\
f&0.85&&1.063&1.8031&2.16&1.25&1.83&1.55&0.362&0.856&1.86&1.61\\
$1^3A_g$&4.12&4.15&3.92&3.47&3.37&3.33&3.34&4.93&4.54&4.87&4.98&4.93\\
$1^3B_u$&2.55&2.4&2.09&1.95&1.49&1.3&1.37&3.13&2.21&2.59&3.04&2.86\\
Octatetraene&&&&&&&&&&&&\\
$2^1A_g$&4.38&4.47&4.84&7.07&6.79&3.17&4.95&4.17&1.57&2.76&4.82&4.23\\
$3^1A_g$&6.56&6.4&6.02&7.5&6.88&5.5&6.58&6.51&4.8&5.08&6.33&6.07\\
$4^1A_g$&7.14&&6.35&7.77&7.69&6.03&7.06&7.05&7.23&6.43&6.96&6.92\\
$1^1B_u$&4.42&4.66&4.02&4.9&4.74&3.76&4.38&4.55&2.5&3.52&4.85&4.49\\
f&1.832&&1.471&2.365&3.06&1.62&2.69&2.21&0.381&1.16&2.73&2.33\\
$2^1B_u$&5.83&5.76&6.78&8.13&7.69&4.4&6.19&6.21&5.82&5.85&6.08&5.31\\
f&0.01&&0.029&0.055&0.031&0.041&0.0026&0.001&1.66&1.02&0.003&0\\
$3^1B_u$&8.44&&7.41&8.69&8.33&7.83&8.18&8.04&7.93&7.93&7.05&6.76\\
f&0.002&&0.145&0.055&0.082&0.129&0.319&&&&0.124&0.362\\
$1^3A_g$&2.17&2.2&1.68&2.89&2.73&2.62&2.7&4.07&3.59&3.99&4.14&4.11\\
$1^3B_u$&3.39&3.55&3.24&1.63&1.09&0.92&1&2.56&1.62&2.08&2.52&2.36\\
Cyclopropene&&&&&&&&&&&&\\
$1^1B_1$&6.36&6.76&6.46&7.4&7.16&7.04&7.16&6.28&6.13&6.27&6.43&6.43\\
f&0.01&&0.001&0.003&0.003&0.003&0.003&0.002&0.002&0.002&0.002&0.002\\
$1^1B_2$&7.45&7.06&6.31&7.01&6.81&6.81&6.81&6.77&6.77&6.77&7.03&7.03\\
f&0.101&&0.074&0.184&0.167&0.167&0.167&0.051&0.052&0.052&0.082&0.082\\
$1^3B_2$&4.18&4.34&3.7&3.26&3.07&3.07&3.07&5.11&5.11&5.11&4.93&4.93\\
$1^3B_1$&6.05&6.62&6.01&6.89&6.68&6.61&6.68&5.91&5.82&5.91&6.06&6.06\\
Cyclopentadiene&&&&&&&&&&&&\\
$2^1A_1$&6.31&6.31&6.52&8.51&8.22&5.93&6.68&6.14&4.52&4.9&6.63&6.28\\
f&0&&0.007&0.02&0.01&0.001&0.001&0.01&0&0&0.013&0.005\\
$3^1A_1$&7.89&&8.15&9.08&8.8&8.8&8.49&9.03&8.11&8.6&9.32&9.21\\
f&0.442&&0.563&1.077&0.981&0.956&0.814&0.488&0.118&0.332&0.754&0.764\\
$1^1B_2$&5.27&5.55&5.02&5.67&5.46&5.21&5.34&5.76&5.22&5.51&5.83&5.77\\
f&0.148&&0.09&0.15&0.157&0.156&0.155&0.142&0.135&0.137&0.148&0.148\\
$1^3A_1$&4.9&5.09&4.75&4.26&4.26&4.26&4.26&5.86&5.86&5.86&5.89&5.89\\
$1^3B_2$&3.15&3.25&2.71&2.4&2.15&2.05&2.09&3.95&3.57&3.76&3.76&3.72\\
Norbornadiene&&&&&&&&&&&&\\
$1^1A_2$&5.28&5.34&4.79&5.8&5.54&5.54&5.54&4.75&4.75&4.75&5.19&5.19\\
$2^1A_2$&7.36&&6.86&8.24&7.93&7.93&7.93&6.81&6.82&6.76&7.28&7.28\\
$1^1B_2$&6.2&6.11&5.52&7.29&7.01&7.01&7.01&5.09&5.09&5.09&5.64&5.64\\
f&0.008&0.029&0.01&0.16&0.124&0.124&0.124&0.006&0.006&0.006&0.009&0.009\\
$2^1B_2$&6.48&&6.87&8.16&7.89&7.89&7.89&7.09&7.09&7.09&7.64&7.64\\
f&0.343&0.187&0.173&0.353&0.338&0.338&0.336&0.063&0.063&0.063&0.124&0.124\\
$1^3A_2$&3.42&3.72&3.08&2.81&2.65&2.65&2.65&4.11&4.11&4.11&4.15&4.15\\
$1^3B_2$&3.8&4.16&3.62&3.16&2.99&2.99&2.99&4.75&4.75&4.75&4.86&4.86\\
\hline
\end{tabular}
\label{tab:results}
\end{table*}
\begin{table*}
\begin{tabular}{ccccccccccccc}
\hline
Benzene&CASPT2&Best&B3LYP&CIS&RI-CIS&D-CIS&x-D-CIS&ALDA&DALDA&x-D-ALDA&hybrid&x-D-hybrid\\
\hline
$1^1B_{1u}$&6.3&6.54&6.1&6.27&6.12&6.12&6.12&6.89&6.89&6.01&7.03&7.03\\
$1^1B_{2u}$&4.84&5.08&5.4&6.44&6.32&6.32&6.32&5.36&5.36&5.36&5.56&5.56\\
$11E_{1u}$&7.03&7.13&7.07&8.29&8.08&8.08&8.08&8.02&8.02&8.02&8.09&8.1\\
f&0.82&&1.195&1.17&1.09&1.09&1.09&0.884&0.884&0.884&0.941&0.941\\
$11E_{2g}$&7.9&8.41&8.91&10.81&10.68&9.18&10.19&9.04&9.04&9.04&9.71&9.7\\
$1^3B_{1u}$&3.89&4.15&3.77&3.34&3.13&3.13&3.13&5.18&5.18&5.18&5.13&5.13\\
$1^3B_{2u}$&5.49&5.88&5.09&5.98&5.86&5.86&5.86&5.34&5.34&5.34&5.38&5.38\\
$13E_{1u}$&4.49&4.86&4.7&5.08&4.92&4.92&4.92&5.27&5.27&5.27&5.27&5.27\\
$13E_{2g}$&7.12&7.51&7.33&7.82&7.69&6.41&6.53&7.96&6.42&7.57&7.73&7.73\\
Naphthalene&&&&&&&&&&&&\\
$2^1A_g$&5.39&5.87&6.18&7.55&7.42&5.14&6.88&5.07&3.75&4.31&5.69&5.4\\
$3^1A_g$&6.04&6.67&6.85&9.13&8.9&6.49&7.17&6.28&6.71&5.82&6.37&6.12\\
$1^1B_{2u}$&4.56&4.77&4.35&5.26&5.06&5.06&5.06&4.47&4.47&4.47&4.83&4.83\\
f&0.05&&0.062&0.114&0.112&0.112&0.122&0.056&0.056&0.056&0.081&0.081\\
$2^1B_{2u}$&5.93&6.33&6.12&7.45&7.22&7.22&7.22&6.49&6.49&6.49&6.83&6.83\\
f&0.313&&0.186&0.684&0.601&0.601&0.601&0.158&0.158&0.158&0.24&0.24\\
$3^1B_{2u}$&7.16&&7.87&9.85&9.67&9.66&9.67&8.29&8.29&8.29&8.62&8.62\\
f&0.848&&0.532&0.806&&&&&&&&\\
$1^1B_{3u}$&4.03&4.24&4.44&5.38&5.16&5.16&5.16&4.3&4.3&4.31&4.56&4.56\\
f&0.001&&0&0&0&0&0&0&0&0&0&0\\
$2^1B_{3u}$&5.54&6.06&5.93&7.23&7.12&7.12&7.11&6.46&6.46&6.46&6.71&6.72\\
f&1.337&&1.268&2.483&2.5&2.5&2.49&1.74&1.74&1.74&2&2\\
$3^1B_{3u}$&7.18&&8.65&&12.21&11.14&9.52&7.71&7.71&7.14&8.67&8.63\\
f&0.048&&0.01&&&&&&&0.033&&\\
$1^1B_{1g}$&5.53&5.99&5.58&6.95&6.78&5.68&6.62&5.95&5.47&6.26&6.37&6.12\\
$2^1B_{1g}$&5.87&6.47&6.32&8.08&7.85&6.65&6.96&7.03&6.71&6.98&7.06&6.44\\
$1^3A_g$&5.27&5.52&5.33&5.41&5.38&5.13&5.27&5.81&5.34&5.05&5.11&5.11\\
$2^3A_g$&5.83&6.47&5.95&7.21&6.95&5.93&6.28&5.92&5.87&5.69&5.67&5.5\\
$3^3A_g$&5.91&6.79&6.07&7.53&7.39&6.7&6.88&6.11&6.16&5.82&6.02&6\\
$1^3B_{2u}$&3.1&3.11&2.69&2.52&2.25&2.25&2.25&3.49&3.49&3.49&3.53&3.53\\
$2^3B_{2u}$&4.3&4.64&4.4&4.84&4.71&4.71&4.71&5.05&5.05&5.05&5.1&5.1\\
$1^3B_{3u}$&3.89&4.18&3.95&4.36&4.23&4.23&4.22&4.18&4.18&4.18&4.35&4.35\\
$2^3B_{3u}$&4.45&5.11&4.22&5.16&4.95&4.95&4.95&4.25&4.25&4.24&4.44&4.44\\
$1^3B_{1g}$&4.23&4.47&4.17&4.02&3.89&3.83&3.86&5.04&4.21&4.61&5.1&5.1\\
$2^3B_{1g}$&5.71&6.48&5.55&7.45&7.17&6.06&6.9&5.2&5.15&5.18&5.67&5.5\\
$3^3B_{1g}$&6.23&6.76&6.56&7.97&7.71&6.64&7.14&6.9&6.37&6.72&7.26&7.19\\
Furan&&&&&&&&&&&&\\
$2^1A_1$&6.16&6.57&6.7&8.25&8.02&5.12&6.97&6.54&3.35&5.37&6.92&6.53\\
f&0.002&&0&0.001&0&0.007&0.028&0&0.001&0.004&0&0.001\\
$3^1A_1$&7.66&8.13&8.25&9.33&9.04&8.41&8.77&9.41&8.45&9.04&9.43&9.28\\
f&0.416&&0.437&0.863&0.756&0.556&0.669&0.514&0.185&0.479&0.607&0.599\\
$1^1B_2$&6.04&6.32&6.16&6.69&6.35&6.12&6.17&7.06&6.51&6.62&7.09&7\\
f&0.154&&0.162&0.216&0.214&0.202&0.213&0.277&0.227&0.238&0.279&0.276\\
$1^3A_1$&5.15&5.48&5.21&4.94&4.97&4.33&4.97&6.1&5.41&6.1&6.1&6.1\\
$1^3B_2$&3.99&4.17&3.71&3.26&2.99&2.84&2.94&4.92&4.39&4.72&4.74&4.69\\
Pyrrole&&&&&&&&&&&&\\
$2^1A_1$&5.92&6.37&6.53&7.79&7.61&6.93&6.96&6.46&6.39&5.54&6.78&6.58\\
f&0.02&&0.001&0.006&0.002&0.009&0.009&0.001&0.001&0.002&0.002&0\\
$3^1A_1$&7.46&7.91&7.96&9.05&8.8&8.59&8.6&8.89&8.46&8.55&8.96&8.91\\
f&0.326&&0.451&0.876&0.78&0.601&0.621&0.383&0.317&0.367&0.628&0.628\\
$1^1B_2$&6&6.57&6.4&6.94&6.7&6.57&6.37&7.27&6.62&6.65&7.27&7.15\\
f&0.125&&0.173&0.236&0.232&0.189&0.198&0.265&0.177&0.183&0.28&0.268\\
$1^3A_1$&5.16&5.51&5.25&5.24&5.24&5.24&5.24&5.91&5.4&5.91&5.93&5.93\\
$1^3B_2$&4.27&4.48&4.07&3.69&3.51&3.44&3.4&5.18&4.73&4.75&5.02&4.91\\
\hline
\end{tabular}
\label{tab:results}
\end{table*}
\begin{table*}
\begin{tabular}{ccccccccccccc}
\hline
Imidazole&CASPT2&Best&B3LYP&CIS&RI-CIS&D-CIS&x-D-CIS&ALDA&DALDA&x-D-ALDA&hybrid&x-D-hybrid\\
\hline
$2^1A'$&6.72&6.19&6.45&7.23&6.97&6.89&6.45&6.72&5.38&5.64&6.97&6.67\\
f&0.126&&0.144&0.26&0.254&0.23&0.211&0.072&0.05&0.061&0.093&0.096\\
$3^1A'$&7.15&6.93&7.04&8.15&7.9&7.3&7.46&7.58&7.05&7.2&7.29&7.25\\
f&0.143&&0.029&0.025&0.05&0.094&0.083&0.137&0.133&0.123&0.004&0.016\\
$4^1A'$&8.51&&8.27&9.43&9.22&9.03&9.85&7.96&7.93&7.96&7.46&7.43\\
f&0.594&&0.359&0.65&0.471&0.328&0.156&0.026&0.014&0.013&0.209&0.192\\
$1^1A"$&6.52&6.81&6.46&7.63&7.5&7.39&7.43&5.56&5.5&5.52&6.14&6.14\\
f&0.011&&0.003&0.014&0.015&0.015&0.015&0.001&0.001&0.001&0.002&0.002\\
$2^1A"$&7.56&&7.45&9.58&9.35&8.47&8.53&7.31&6.92&7.03&6.37&6.36\\
f&0.013&&0.005&0&0.001&0.001&0.001&0.02&0.018&0.019&0.001&0\\
$1^3A'$&4.49&4.69&4.24&3.9&3.72&3.72&3.71&5.29&5.23&5.23&5.12&5.11\\
$2^3A'$&5.47&5.79&5.44&5.38&5.32&5.32&5.29&6.3&6.23&6.3&6.19&6.18\\
$3^3A'$&6.53&6.55&5.95&6.59&6.22&6.22&6.22&6.6&6.5&6.51&6.56&6.55\\
$4^3A'$&7.08&&6.93&7.92&7.58&7.58&7.52&7.64&7.64&7.64&7.54&7.54\\
$1^3A"$&6.07&6.37&5.83&6.39&6.3&6.22&6.24&5.33&5.3&5.31&5.78&5.78\\
$2^3A"$&7.15&&6.86&7.72&8.72&7.83&7.78&&&&6.59&6.51\\
Pyridine&&&&&&&&&&&&\\
$2^1A_1$&6.42&6.26&6.31&6.69&6.55&6.45&6.54&7.07&6.87&7.07&7.23&7.23\\
f&0.005&&0.016&0.01&0.011&0.002&0.095&0.014&0.002&0.012&0.025&0.024\\
$3^1A_1$&7.23&7.18&7.32&8.59&8.37&7.46&8.35&8.32&8.37&8.31&8.43&8.43\\
f&0.82&&0.424&1.049&0.956&0.138&0.955&0.604&0.543&0.643&0.774&0.775\\
$1^1B_2$&4.84&4.85&5.49&6.39&6.19&6.17&6.18&5.51&5.49&5.51&5.71&5.71\\
f&0.018&&0.035&0.064&0.078&0.071&0.077&0.018&0.016&0.018&0.023&0.023\\
$2^1B_2$&7.48&7.27&7.3&8.58&8.4&7.75&8.39&8.05&7.39&8.05&8.13&8.13\\
f&0.64&&0.455&0.95&0.843&0.151&0.843&0.654&0.002&0.659&0.515&0.517\\
$1^1B_1$&4.91&4.59&4.8&5.89&5.69&5.5&5.55&4.29&3.91&2.99&4.8&4.8\\
f&0.009&&0.004&0.011&0.011&0.01&0.011&0.005&0.003&0.002&0.007&0\\
$1^1A_2$&5.17&5.11&5.11&7.38&7.23&7.22&7.23&4.12&4.11&4.11&4.77&4.39\\
$1^3A_1$&4.05&4.06&3.89&3.42&3.17&3.17&3.17&5.36&5.36&5.36&5.22&5.22\\
$2^3A_1$&4.73&4.91&4.84&5.18&5.01&5.01&5.01&5.6&5.6&5.6&5.49&5.49\\
$3^3A_1$&7.34&&7.44&7.82&7.66&7.66&7.66&8.16&8.16&8.16&8.38&8.38\\
$1^3B_2$&4.56&4.64&4.51&5.01&4.76&4.76&4.76&4.93&4.93&4.93&5.01&5.01\\
$2^3B_2$&6.02&6.08&5.64&6.46&6.35&6.31&6.35&5.87&5.81&5.86&6&6\\
$3^3B_2$&7.28&&7.75&8.33&8.23&7.78&8.23&8.5&8.5&8.5&8.7&8.7\\
$1^3B_1$&4.41&5.25&4.04&4.77&4.62&4.47&4.5&3.71&3.45&3&4.07&3.87\\
$1^3A_2$&5.1&5.28&4.98&7.17&7.02&7.01&7.02&4.02&4.01&4.02&4.69&4.69\\
Pyrazine&&&&&&&&&&&&\\
$1^1B_{1u}$&6.7&6.58&6.5&6.86&6.72&6.43&6.65&7.38&7.53&7.32&7.48&7.48\\
f&0.08&&0.059&0.039&0.042&0.001&0.025&0.071&0.157&0.037&0.101&0.096\\
$2^1B_{1u}$&7.57&7.72&7.68&8.9&8.68&7.36&8.4&8.31&8.39&8.27&8.64&8.62\\
f&0.76&&0.367&0.903&0.795&0.193&0.705&0.039&0&0.05&0.375&0.388\\
$1^1B_{2u}$&4.75&4.64&5.37&6.25&5.93&5.91&5.92&5.52&5.48&5.51&5.71&5.7\\
f&0.07&&0.091&0.171&0.182&0.127&0.18&0.062&0.054&0.059&0.077&0.076\\
$2^1B_{2u}$&7.7&7.6&7.78&9.13&9.07&7.84&9&8.47&8.7&8.42&8.67&8.66\\
f&0.66&&0.264&0.73&0.597&0.108&0.589&0.689&0.69&0.691&0.819&0.819\\
$1^1A_{u}$&4.52&4.81&4.69&6.84&6.61&6.61&6.61&3.62&3.62&3.62&4.27&4.27\\
$1^1B_{1g}$&6.13&6.6&6.38&9.75&9.62&9.62&9.62&5.17&5.17&5.17&6.05&6.05\\
$1^1B_{2g}$&5.17&5.56&5.55&6.53&6.34&6.34&6.34&5.04&5.04&5.04&5.66&5.66\\
$1^1B_{3u}$&3.63&3.95&3.96&4.9&4.63&4.63&4.63&3.44&3.44&3.44&3.88&3.88\\
f&0.01&&0.006&0.016&0.016&0.016&0.0157&0.007&0.007&0.007&0.009&0.009\\
Pyrimidine&&&&&&&&&&&&\\
$2^1A_1$&6.72&6.95&6.58&7.04&6.87&6.9&6.84&7.32&7.26&7.3&7.52&7.51\\
f&0.05&&0.037&0.021&0.025&0.04&0.023&0.059&0.037&0.051&0.074&0.07\\
$3^1A_1$&7.57&&7.48&8.75&8.55&8.62&8.53&8.4&7.75&8.36&8.34&8.32\\
f&0.58&&0.386&0.863&0.778&0.72&0.777&0.498&0.055&0.5&0.317&0.325\\
$1^1B_2$&4.93&5.44&5.74&6.69&6.48&6.43&6.49&5.74&5.7&5.73&5.95&5.95\\
f&0.001&&0.034&0.068&0.081&0.077&0.081&0.018&0.018&0.018&0.023&0.023\\
$2^1B_2$&7.32&&7.76&8.99&8.84&8.89&8.82&8.58&8.62&8.57&7.6&7.6\\
f&0.79&&0.297&0.852&0.764&0.745&0.759&0.411&0.483&0.365&0.007&0.007\\
$1^1B_1$&3.81&4.55&4.27&5.64&5.41&5.41&5.41&3.59&3.59&3.59&4.14&4.14\\
f&0.02&&0.005&0.018&0.019&0.019&0.019&0.005&0.005&0.005&0.008&0.075\\
$1^1A_2$&4.12&4.91&4.6&6.31&6.13&6.13&6.13&3.68&3.68&3.68&4.33&4.33\\
\hline
\end{tabular}
\label{tab:results}
\end{table*}
\begin{table*}
\begin{tabular}{ccccccccccccc}
\hline
Pyridazine&CASPT2&Best&B3LYP&CIS&RI-CIS&D-CIS&x-D-CIS&ALDA&DALDA&x-D-ALDA&hybrid&x-D-hybrid\\
\hline
$2^1A_1$&4.86&5.18&5.61&6.52&6.32&6.42&6.3&5.62&4.48&5.6&5.83&5.82\\
f&0.009&&0.022&0.044&0.051&0.045&0.053&0.009&0.003&0.01&0.013&0.013\\
$3^1A_1$&7.5&&7.5&8.8&8.6&7.3&8.59&8.45&8.48&8.42&8.13&8.13\\
f&0.5&&0.335&0.873&0.766&0.062&0.706&0.489&0.412&0.513&0.572&0.572\\
$1^1B_2$&6.61&&6.43&6.76&6.62&6.35&6.57&7.11&7.2&7.09&6.75&6.75\\
f&0.003&&0.002&0.002&0.004&0.002&0.005&0.002&0&0.0039&0.002&0.002\\
$2^1B_2$&7.39&&7.24&8.47&8.22&8.3&8.2&8.07&8.1&8.04&7.36&7.35\\
f&0.75&&0.431&0.855&0.756&0.72&0.75&0.535&0.506&0.539&0.001&0.002\\
$1^1A_2$&3.66&4.31&4.18&5.83&5.65&5.65&5.6&3.16&3.16&3.16&3.87&3.87\\
$2^1A_2$&5.09&5.77&5.44&7.18&6.84&6.84&6.68&4.87&4.87&4.87&5.35&5.35\\
$1^1B_1$&3.48&3.78&3.58&4.71&4.45&4.45&4.22&2.99&2.99&2.44&3.5&3.34\\
f&0.01&&0.005&0.017&0.017&0.017&0.015&0.005&0.005&0.003&0.007&0.007\\
$2^1B_1$&5.8&&6.09&8.3&8.12&8.12&8.01&5.1&5.1&5.06&5.82&5.82\\
f&0.008&&0.005&0.007&0.008&0.008&0.001&0.005&0.005&0.003&0.013&0.013\\
\textit{s}-triazine&&&&&&&&&&&&\\
$2^1A'$&6.77&&7.01&7.52&7.31&6.88&7.3&7.84&7.92&7.82&7.24&7.24\\
$1^1A_2'$&5.53&5.79&6.14&7.2&7.08&6.79&7.08&6.02&5.96&6.01&6.25&6.25\\
$11E'$&8.16&&7.79&9.1&8.93&8.96&8.91&8.49&8.53&8.48&8.66&8.65\\
f&0.61&&0.762&0.768&0.717&0.704&0.716&0.258&0.246&&0.463&0.468\\
$1^1A_1"$&3.9&4.6&4.45&6.6&6.51&6.51&6.51&3.39&3.39&3.39&4.09&4.06\\
$1^1A_2"$&4.08&4.66&4.54&5.93&5.68&5.68&5.68&3.9&3.9&3.9&4.41&4.41\\
f&0.015&&0.014&0.039&0.042&0.042&0.042&0.018&0.018&&0.024&0.024\\
$11E"$&4.36&4.7&4.54&6.16&5.98&5.98&5.98&3.64&3.64&3.64&4.26&4.26\\
$21E"$&7.15&&7.49&9.44&9.32&9.32&9.32&7.64&7.64&7.64&7.26&7.26\\
\textit{s}-tetrazine&&&&&&&&&&&&\\
$1^1A_{u}$&3.06&3.51&3.51&5.27&5.02&4.22&5.02&2.38&2.38&2.38&3.17&3.17\\
$2^1A_{u}$&5.28&5.5&5.04&6.44&6.03&6.35&6.03&4.22&4.22&4.22&4.81&4.81\\
$1^1B_{1g}$&4.51&4.73&4.73&5.93&5.71&5.71&5.71&3.76&3.76&3.76&4.67&4.67\\
$2^1B_{1g}$&5.99&&6.64&9.76&9.47&9.47&9.47&5.35&5.35&5.35&6.49&6.49\\
$3^1B_{1g}$&6.2&&7.4&11.96&11.48&11.23&11.48&6.38&6.38&6.38&7.04&7.04\\
$1^1B_{2g}$&5.05&5.2&5.29&6.44&6.11&6.11&6.11&4.39&4.39&4.39&5.21&5.21\\
$2^1B_{2g}$&5.48&&5.99&9.32&9.06&9.06&9.06&5.04&5.04&5.04&5.84&5.84\\
$2^1B_{3g}$&8.12&&9.3&10.53&10.77&10.05&10.77&8.1&8.1&8.1&8.58&8.58\\
$1^1B_{1u}$&7.13&&6.9&7.08&6.92&7.04&6.85&7.81&7.84&7.71&8.05&8.04\\
f&0.001&&0.002&0&0&0.002&0&0.033&0.059&0.023&0.004&0.004\\
$2^1B_{1u}$&7.54&&7.48&8.7&8.43&8.48&8.43&8.24&8.32&8.23&8.42&8.42\\
f&0.687&&0.337&0.645&0.559&0.546&0.559&0.345&0.307&0.355&0.436&0.436\\
$1^1B_{2u}$&4.89&4.93&5.58&6.51&6.22&6.35&6.11&5.71&5.82&5.64&5.92&5.91\\
f&0.045&&0.064&0.133&0.14&0.126&0.139&0.039&0.032&0.042&0.054&0.055\\
$2^1B_{2u}$&7.94&&8.26&9.53&9.42&9.43&9.46&9&9.02&9.13&9.19&9.19\\
f&0.733&&0.29&0.562&0.462&0.464&0.459&0.361&&&0.445&0.444\\
$1^1B_{3u}$&1.96&2.29&2.24&3.33&2.94&2.94&2.94&1.62&1.62&1.62&2.11&2.11\\
f&0.013&&0.005&0.022&0.021&0.021&0.021&0.006&0.006&0.006&0.01&0.01\\
$2^1B_{3u}$&6.37&&6.29&8.34&8.14&8.14&8.14&5.08&5.08&5.08&5.93&5.93\\
f&0.017&&0.01&0.023&0.026&0.026&0.026&0.011&0.011&0.011&0.016&0.016\\
$1^3A_{u}$&2.81&3.52&3.1&4.23&4.04&4.02&4.04&2.15&2.15&2.15&2.86&2.86\\
$2^3A_{u}$&4.85&5.03&4.43&6.07&5.64&4.36&5.64&3.69&3.69&3.69&4.23&4.23\\
$1^3B_{1g}$&3.76&4.21&3.63&4.13&3.98&3.98&3.98&3.16&3.16&3.16&3.69&3.69\\
$2^3B_{1g}$&5.68&&6.33&9.67&9.37&9.37&9.37&5.32&5.32&5.32&6.13&6.13\\
$1^3B_{1u}$&4.25&4.33&3.83&3.04&2.74&2.74&2.74&5.7&5.7&5.66&5.44&5.44\\
$2^3B_{1u}$&5.09&5.38&5.24&5.69&5.55&5.56&5.5&5.96&5.96&5.94&5.94&5.93\\
$1^3B_{2g}$&4.67&4.93&4.48&5.13&4.91&4.91&4.91&4.17&4.17&4.17&4.61&4.61\\
$2^3B_{2g}$&5.3&&5.62&8.96&8.66&8.66&8.66&4.33&4.33&4.33&5.33&5.33\\
$1^3B_{2u}$&4.29&4.54&4.06&4.41&4.02&4.02&4.02&4.68&4.68&4.68&4.67&4.67\\
$2^3B_{2u}$&6.81&&6.63&7.59&7.6&7.6&7.6&6.8&6.8&6.8&7.02&7.02\\
$1^3B_{3u}$&1.45&1.89&1.42&2.07&1.74&1.74&1.74&0.99&0.99&0.99&1.36&1.36\\
$2^3B_{3u}$&6.14&&5.97&7.9&7.7&7.7&7.7&4.85&4.85&4.85&5.64&5.64\\
\hline
\end{tabular}
\label{tab:results}
\end{table*}
\begin{table*}
\begin{tabular}{ccccccccccccc}
\hline
Formaldehyde&CASPT2&Best&B3LYP&CIS&RI-CIS&D-CIS&x-D-CIS&ALDA&DALDA&x-D-ALDA&hybrid&x-D-hybrid\\
\hline
$1^1A_2$&3.91&3.88&3.89&4.18&4.18&2.84&3.12&3.87&2.34&2.63&3.97&3.59\\
$1^1B_1$&9.09&9.1&8.89&9.19&9.21&9.2&9.2&9.02&9.01&9.02&9.06&9.06\\
f&0.01&&0.001&0.002&0.001&0.001&0.001&0.003&0.003&0.003&0.003&0.003\\
$2^1A_1$&10.08&9.3&9.17&9.7&9.62&9.71&9.42&11.67&11.4&9.66&11.63&11.62\\
f&0.28&&0.35&0.259&0.206&0.203&0.175&0.241&0.209&0.03&0.404&0.403\\
$1^3A_2$&3.48&3.5&3.13&3.4&3.4&2.65&2.82&3.23&2.34&2.52&3.33&3.1\\
$1^3A_1$&5.99&5.87&5.18&4.27&4.08&4.08&4.08&7.55&7.55&7.55&6.96&6.96\\
Acetone&&&&&&&&&&&&\\
$1^1A_2$&4.18&4.4&4.34&4.88&4.77&3.69&4.1&4.28&3.15&3.53&4.38&4.1\\
$1^1B_1$&9.1&9.17&8.6&9.4&9.35&9.36&9.29&7.83&7.82&7.83&8.49&8.49\\
f&0.01&&0&0&0&0&0&0.004&0.004&0.004&0.002&0.002\\
$2^1A_1$&9.16&9.65&9.04&9.67&9.62&9.73&9.53&9.02&9.02&9.02&9.75&9.74\\
f&0.326&&0.195&0.371&0.316&0.292&0.309&0.137&0.084&0.136&0.209&0.21\\
$1^3A_2$&3.9&4.05&3.69&4.17&4.07&3.49&3.71&3.74&3.13&3.32&3.83&3.66\\
$1^3A_1$&5.98&6.03&5.39&4.8&4.62&4.62&4.62&6.86&6.86&6.86&6.73&6.73\\
\textit{o}-benzoquinone&&&&&&&&&&&&\\
$1^1A_{u}$&2.5&2.77&2.58&3.92&3.58&1.12&2.63&2.05&-0.74&0.58&2.62&2.04\\
$1^1B_{1g}$&2.5&2.76&2.43&3.73&3.32&3.32&3.32&1.83&1.83&1.83&2.33&2.33\\
$1^1B_{1u}$&5.15&5.28&4.83&6.23&6.13&6.18&6.01&4.93&4.98&4.91&5.37&5.37\\
f&0.616&&0.323&1.154&1.23&1.2&1.14&0.242&0.225&0.284&0.36&0.362\\
$2^1B_{1u}$&7.08&&7.25&8.89&8.5&8.5&8.5&7.38&7.4&7.2&8.09&8.08\\
f&0.624&&0.561&0.721&0.724&0.725&0.73&0.231&0.225&0.192&0.487&0.487\\
$1^1B_{3g}$&4.19&4.26&3.73&5.21&4.51&4.51&4.51&3.31&3.31&3.31&3.77&3.77\\
f&0&&0&0&0&0&0&0&0&0&0&0\\
$2^1B_{3g}$&6.34&6.96&6.59&8.58&8.46&8.46&8.46&6.46&6.54&6.54&7.06&7.06\\
$1^1B_{3u}$&5.15&5.64&5.43&8.27&7.96&7.97&7.97&4.41&4.41&4.41&5.41&5.41\\
f&0&&0&0.016&0.014&0.014&0.014&0&0&0&0&0\\
$1^3A_{u}$&2.27&2.62&2.05&3.22&2.88&1.22&2.29&1.73&-0.51&0.6&2.18&1.91\\
$1^3B_{1g}$&2.17&2.51&1.92&3.05&2.67&2.67&2.67&1.48&1.48&1.48&1.91&1.78\\
$1^3B_{1u}$&2.91&2.96&2.19&2.04&1.42&1.42&1.42&3.15&3.13&3.15&3.09&3.09\\
$1^3B_{3g}$&3.19&3.41&2.68&2.72&2.36&2.36&2.35&2.89&2.89&2.89&3.13&3.13\\
Formamide&&&&&&&&&&&&\\
$2^1A'$&7.41&7.39&8.13&8.73&8.67&7.47&7.93&8.82&8.92&7.98&8.78&8.76\\
f&0.371&&0.371&0.278&0.206&0.078&0.032&0.383&0.036&0.346&0.123&0.179\\
$3^1A'$&10.5&&10.92&10.55&10.63&9.08&9.52&12.05&10.29&10.4&9.17&8.86\\
f&0.131&&0.055&0.193&0.343&0.394&0.102&0.055&0.081&0.079&0.542&0.453\\
$1A"$&5.61&5.63&5.55&6.13&6.13&4.92&5.27&5.56&4.21&4.57&5.72&5.4\\
f&0.001&&0.001&0.002&0.003&0.001&0.002&0.002&0&0.001&0.002&0.002\\
$1^3A'$&5.69&5.74&5.13&5.14&4.86&4.85&4.86&5.93&5.93&4.63&5.92&5.92\\
$1^3A"$&5.34&5.36&4.97&5.51&5.47&4.69&4.93&5.06&4.21&4.42&5.21&4.99\\
Acetamide&&&&&&&&&&&&\\
$2^1A'$&7.21&7.27&7.46&8.9&8.95&8.95&9.02&8.1&8.89&7.07&7.9&7.77\\
f&0.292&&0.087&0.248&0.206&0.296&0.324&0.161&0.166&0.131&0.062&0.122\\
$3^1A'$&10.08&&10.01&11.51&11.2&10.25&9.91&9.65&9.52&9.14&8.45&8.28\\
f&0.179&&0.224&0.114&0.162&0.034&0.202&0.173&0.05&0.142&0.156&0.09\\
$1^1A"$&5.54&5.69&5.56&6.36&6.3&5.31&5.54&5.51&4.46&4.63&5.65&5.38\\
f&0.001&&0.001&0.001&0.001&0.001&0.001&0.002&0.001&0.001&0.003&0.002\\
$1^3A'$&5.57&5.88&5.26&5.41&5.16&5.16&5.16&5.91&5.91&4.6&5.95&5.95\\
$1^3A"$&5.24&5.42&5.01&5.73&5.65&5.08&5.2&5.03&4.38&4.48&5.17&4.99\\
Propanamide&&&&&&&&&&&&\\
$2^1A'$&7.28&7.2&7.76&8.92&8.92&7.42&8.32&7.77&7.77&7.77&6.95&6.29\\
f&0.346&&0.107&0.287&0.206&0.121&0.038&0.067&0.067&0.068&0.012&0.011\\
$3^1A'$&9.95&&9&10.06&9.79&9.08&9.01&8.11&7.99&7.86&8.01&7.79\\
f&0.205&&0.085&0.091&0.106&0.216&0.278&0.121&0.072&0.084&0.085&0.154\\
$1^1A"$&5.48&5.72&5.59&6.34&6.31&5.41&5.63&5.52&4.56&4.76&5.7&5.46\\
f&0.001&&0&0.001&0.001&0.001&0.001&0.001&0.001&0.001&0.001&0.001\\
$1^3A'$&5.94&5.9&5.28&5.46&5.22&5.22&5.22&5.89&5.89&5.89&5.97&5.97\\
$1^3A"$&5.28&5.45&5.04&5.76&5.67&5.16&5.27&5.05&4.45&4.58&5.21&5.06\\
\hline
\end{tabular}
\label{tab:results}
\end{table*}
\begin{table*}
\begin{tabular}{ccccccccccccc}
\hline
Cytosine&CASPT2&Best&B3LYP&CIS&RI-CIS&D-CIS&x-D-CIS&ALDA&DALDA&x-D-ALDA&hybrid&x-D-hybrid\\
\hline
$2^1A'$&4.39&4.66&4.64&6.09&5.94&5.18&5.64&4.51&3.47&3.78&4.99&4.33\\
f&0.061&&0.035&0.161&0.182&0.069&0.105&0.022&0.014&0.005&0.046&0.014\\
$3^1A'$&5.36&5.62&5.42&7.42&7.28&6.38&7.03&5.11&4.9&4.91&5.73&5.55\\
f&0.108&&0.087&0.361&0.12&0.19&0.264&0.042&0.051&0.02&0.069&0.049\\
$4^1A'$&6.16&&6.72&7.91&7.83&7.03&7.79&6.17&5.98&5.89&6.76&6.41\\
f&0.863&&0.368&0.819&1.03&0.176&0.934&0.115&0.019&0.124&0.136&0.196\\
$5^1A'$&6.74&&6.46&8.98&8.75&7.76&8.43&6.98&6.9&6.04&7.25&7.02\\
f&0.147&&0.177&0.186&0.364&0.416&0.233&0.539&0.43&0.074&0.505&0.415\\
$1^1A"$&5&4.87&4.76&6.6&6.44&5.27&6.44&3.82&3.73&3.78&4.68&4.62\\
f&0.005&&0.001&0.003&0.003&0.002&0.04&0&0&0&0.001&0.009\\
$2^1A"$&6.53&5.26&5.11&6.91&6.93&6.76&6.93&4.27&3.05&4.23&4.5&5.10\\
f&0.001&&0.001&0.001&0.001&0&0.001&0.002&0.001&0.001&0&0.001\\
Thymine&&&&&&&&&&&&\\
$2^1A'$&4.88&5.2&5&6.36&6.1&5.92&6.02&4.79&4.1&4.61&5.35&5.16\\
f&0.17&&0.136&0.497&0.463&0.456&0.429&0.064&0.017&0.055&0.128&0.088\\
$3^1A'$&5.88&6.27&5.97&8.16&7.8&6.12&7.3&5.62&4.56&5.81&6.31&5.88\\
f&0.17&&0.071&0.202&0.29&0.018&0.288&0.099&0.069&0.073&0.14&0.117\\
$4^1A'$&6.1&6.53&6.31&8.55&8.47&7.41&8.33&6.05&5.54&6.4&6.66&6.38\\
f&0.15&&0.142&0.446&0.271&0.017&0.243&0.085&0.038&0.023&0.139&0.068\\
$5^1A'$&7.16&&7.47&9.48&9.44&7.88&9.21&7.13&6.77&7.16&7.92&7.39\\
f&0.85&&0.411&0.182&0.173&0.198&0.408&0.11&0.059&0.092&0.542&0.189\\
$1^1A"$&4.39&4.82&4.7&6.01&5.96&4.97&5.96&4.21&2.98&3.66&4.81&4.24\\
$2^1A"$&5.91&6.16&5.8&7.41&7.12&7.12&7.05&4.83&4.45&4.69&5.89&5.83\\
$3^1A"$&6.15&&6.21&7.65&7.43&7.37&7.43&5.21&5.21&5.19&6.18&6.09\\
$4^1A"$&6.7&&6.69&8.83&8.24&8.18&8.24&6.13&6.06&5.21&6.7&6.66\\
f&0&&0&0&0&0&0&0&0&0&0&0\\
Uracil&&&&&&&&&&&&\\
$2^1A'$&5&5.35&5.19&6.55&6.31&5.82&6.31&4.95&4.5&4.95&5.52&5.51\\
f&0.19&&0.13&0.51&0.475&0.323&0.475&0.044&0.011&0.044&0.105&0.105\\
$3^1A'$&5.82&6.26&5.87&8.29&8&6.5&8&5.51&5.32&5.51&6.2&6.19\\
f&0.08&&0.04&0.154&0.219&0.096&0.217&0&0.073&0.047&0.066&0.067\\
$4^1A'$&6.46&6.7&6.5&8.66&8.51&7.63&8.51&6.43&6.19&6.18&6.84&6.83\\
f&0.29&&0.12&0.43&0.344&0.011&0.345&0.112&0.035&0.036&0.145&0.151\\
$5^1A'$&7&&7.45&9.35&9.33&8.1&9.32&7.45&6.41&6.43&7.32&7.42\\
f&0.76&&0.44&0.266&0.425&0.267&0.423&0.237&0.078&0.111&0.028&0.032\\
$1^1A"$&4.54&4.8&4.63&6&5.95&4.57&5.62&4.09&2.28&3.46&4.74&4.09\\
$2^1A"$&6&6.1&5.74&7.35&7.33&7.32&7.33&4.79&4.35&4.66&5.64&5.78\\
f&0&&0&0&0.001&0&0.001&0&0&0&0&0\\
$3^1A"$&6.37&6.56&6.14&7.84&7.37&7.34&7.37&5.12&5.11&5.12&6.11&5.99\\
$4^1A"$&6.95&&6.64&9.01&8.47&7.77&8.47&6.09&6.05&6.1&6.65&6.63\\
f&0&&0&0.019&0.017&0&0.017&0&0&0&0&0\\
Adenine&&&&&&&&&&&&\\
$2^1A'$&5.13&5.25&5.27&6.32&6.04&5.66&5.86&4.7&3.69&4.06&5.27&4.71\\
f&0.07&&0.047&0.418&0.459&0.062&0.377&0.083&0.034&0.059&0.11&0.080\\
$3^1A'$&5.2&5.25&5&6.51&6.27&6.08&6.13&5.21&5&4.9&5.6&5.41\\
f&0.37&&0.195&0.041&0.025&0.367&0.097&0.072&0.013&0.016&0.182&0.001\\
$4^1A'$&6.24&&6.32&7.81&7.55&6.54&7.33&5.69&5.33&5.46&6.36&5.95\\
f&0.851&&0.24&0.353&0.182&0.054&0.05&0.077&0.049&0.143&0.032&0.117\\
$5^1A'$&6.72&&6.69&8.28&8.04&6.94&7.85&5.96&5.47&5.72&6.63&6.46\\
f&0.159&&0.107&0.48&0.704&0.023&0.744&0.065&0.126&0.0001&0.201&0.011\\
$6^1A'$&6.99&&7.08&8.45&8.29&7.7&8.13&6.46&6.05&6.07&7.15&6.68\\
f&0.565&&0.137&0.571&0.492&0.589&0.324&0.034&0.002&0.09&0.218&0.264\\
$7^1A'$&7.57&&7.52&9.08&8.84&8.00&8.61&6.68&6.08&6.08&7.58&7.34\\
f&0.406&&0.244&0.26&0.208&0.060&0.325&0.055&0.023&0.026&0.127&0.127\\
$1^1A"$&6.15&5.12&4.97&6.84&6.64&6.27&6.64&3.95&3.61&3.95&4.7&4.71\\
f&0.001&&0&0&0.001&0.006&0.001&0&0&0&0&0.008\\
$2^1A"$&6.86&5.75&5.61&7.25&7.01&6.85&7.01&4.78&4.4&4.72&5.42&5.41\\
f&0.001&&0.001&0.002&0.003&0.003&0.003&0&0.001&0&0.001&0.001\\
\hline
\end{tabular}
\end{table*}
\clearpage